\documentclass[12pt]{article}
\usepackage[dvips]{graphics}
\usepackage{amssymb}
\newcommand{\nc}{\newcommand}
\nc{\figcap}[1]{\begin{quote}\refstepcounter{figure}

        {\bf Figure \thefigure}: {\small #1}\end{quote}}
\nc{\fig}[1]{\mbox{Fig.~\ref{#1}}}
\nc{\bea}{\begin{eqnarray}}
\nc{\eea}{\end{eqnarray}}
\nc{\bean}{\begin{eqnarray*}}
\nc{\eean}{\end{eqnarray*}}
\nc{\ba}{\begin{array}}
\nc{\ea}{\end{array}}
\nc{\be}{\begin{equation}}
\nc{\ee}{\end{equation}}
\nc{\nn}{\nonumber}
\nc{\bra}[1]{\langle #1|}
\nc{\ket}[1]{|#1\rangle}
\nc{\av}[1] {\langle #1\rangle}
\nc{\vac}[1] {\langle 0| #1|0\rangle}
\nc{\amp}[2]{\langle #1|#2\rangle}
\nc{\da}{\dagger}
\nc{\pa}{\partial}
\nc{\ga}{\gamma}
\nc{\ep}{\epsilon}
\nc{\tf}{t_f}
\nc{\half}{\ensuremath{\frac{1}{2}}}
\nc{\hHH}{\hat H}
\nc{\ha}{\hat a}
\nc{\hO}{\hat O}
\nc{\hAA}{\hat A}
\nc{\hB}{\hat B}
\nc{\hG}{\hat G}
\nc{\hN}{\hat N}
\nc{\hU}{\hat U}
\nc{\hx}{\hat{x}}
\nc{\hp}{\hat{p}}
\nc{\hpsi}{\hat \psi}
\nc{\hphi}{\hat \phi}
\nc{\hpi}{\hat \pi}
\nc{\hpd}{\hat \psi ^\dagger}
\nc{\hE}{\hat E}
\nc{\hb}{\hat b}
\nc{\hc}{\hat c}
\nc{\hjo}{\hat j _0}
\nc{\hrho}{\hat \rho}
\nc{\leave}{\! \! \! \! \! / \, \,}
\nc{\intl}[1]{\int d\! #1 \,} %defines a good spacing for integrals
\nc{\intll}[3]{\int _#1^#2 d\! #3 \,} % integral with limits
\nc{\lm}{\lim _{y \rightarrow x}}

\nc{\scd}{\partial ^2 _{A_T}}
\nc{\fd}[1]{\frac{\delta }{\delta #1}} % functional derivative
\nc{\pad}[1]{\frac{\partial}{\partial #1}} % partial derivative
\nc{\refpa}[1]{(\ref{#1})} % referencing equations with brackets

\nc{\calH}{\ensuremath{\mathcal{H}}}
\nc{\calD}{\ensuremath{\mathcal{D}}}
\nc{\calL}{\ensuremath{\mathcal{L}}}
\nc{\calO}{\ensuremath{\mathcal{O}}}
\nc{\hcalO}{\ensuremath{\hat \mathcal{O}}}
\nc{\calK}{\ensuremath{\mathcal{K}}}
\nc{\Tr}{\ensuremath{\mathrm{Tr}}}
\nc{\tr}{\ensuremath{\mathrm{tr}}}
\nc{\ra}{\rightarrow}
\nc{\lr}{\leftrightarrow}
\nc{\phistar}{\phi^*}
\nc{\etat}{\eta_T}
\nc{\het}{\hat E_T}
\nc{\hpt}{\hat \psi_T}
\nc{\hpdt}{\hat \psi ^\dagger_T}
\nc{\bart}{\bar{t}}
\nc{\barp}{\bar{p}}
\nc{\barT}{\bar{T}}
\nc{\hbarrho}{\hat{\bar{\rho}}}
\nc{\bga}{\ensuremath{\mbox{\boldmath{$\gamma$}}}}
\nc{\bsi}{\ensuremath{\mathbf{\sigma}}}
\nc{\bx}{\ensuremath{\mathbf{x}}}
\nc{\by}{\ensuremath{\mathbf{y}}}
\nc{\bz}{\ensuremath{\mathbf{z}}}
\nc{\bp}{\ensuremath{\mathbf{p}}}
\nc{\bn}{\ensuremath{\mathbf{n}}}
\nc{\bbp}{\ensuremath{\bar{\mathbf{p}}}}
\nc{\bP}{\ensuremath{\mathbf{P}}}
\nc{\hbA}{\hat{\ensuremath{\mathbf{A}}}}
\nc{\hbB}{\hat{\ensuremath{\mathbf{B}}}}
\nc{\bA}{\ensuremath{\mathbf{A}}}
\nc{\bJ}{\ensuremath{\mathbf{J}}}
\nc{\bB}{\ensuremath{\mathbf{B}}}
\nc{\bH}{\ensuremath{\mathbf{H}}}
\nc{\bM}{\ensuremath{\mathbf{M}}}
\nc{\bD}{\ensuremath{\mathbf{D}}}
\nc{\bE}{\ensuremath{\mathbf{E}}}
\nc{\hbE}{\hat{\ensuremath{\mathbf{E}}}}
\nc{\br}{\ensuremath{\mathbf{r}}}
\nc{\bj}{\ensuremath{\mathbf{j}}}
\nc{\bOm}{\ensuremath{\mathbf{\Om}}}
\nc{\om}{\omega}
\nc{\Om}{\Omega}
\nc{\sgn}{\mbox{sgn}}
\nc{\deltabar}{\mbox{$\delta\hspace*{-8pt}\vspace*{-8pt}-$}}
\nc{\gammat}{\tilde{\gamma}}
\nc{\binom}[2] {{#1\choose #2}}
\nc{\mub}{\bar{\mu}}
\nc{\rhob}{\bar{\rho}}
\nc{\Bb}{\bar{B}}
\nc{\Jb}{\bar{J}}
\nc{\Mb}{\bar{M}}
\nc{\Tb}{\bar{T}}
\nc{\sbar}{\bar{s}}
\nc{\betab}{\bar{\beta}}
\nc{\hj}{\hat j}
\nc{\hQ}{\hat Q}
\nc{\hJ}{\hat J}
\nc{\hA}{\hat A}
\nc{\hH}{\hat H}
\nc{\de}{\delta}
\nc{\leri}{\leftrightarrow}
\nc{\llabel}[1]{\label{#1}\marginpar{#1}}
\nc{\bc}{\begin{center}}
\nc{\ec}{\end{center}}
\nc{\inv}[1]{\frac{1}{#1}}
\newlength{\overeqskip}
\newlength{\undereqskip}
\nc{\eq}[1]{\mbox{Eq.~(\ref{#1})}}
\nc{\eps}{\epsilon}
\nc{\goto}{\rightarrow}
\nc{\cF}{{\cal F}}
\nc{\cG}{{\cal G}}
\nc{\cH}{{\cal H}}
\begin{document}

%%%%%%%%%%%%%%%%%%%%%%%%%%%%%%%%%   first_page.tex (start)  %%%%%%%%%%%%%%%%%%%%%%%%%%%%%%%%%
% 
\thispagestyle{empty} 
\begin{flushright}{\begin{tabular}{l} 
Revised Version 1999\\ 
Theoretical Physics Seminar\\ 
 in Trondheim No 6 1998 
\end{tabular}} 
\end{flushright} 
\vspace{10mm} 
\begin{center} 
\baselineskip 1.2cm 
{\Huge\bf   Quantum Dynamics of Non-Degenerate Parametric Amplification 
}\\[1mm] 
\normalsize 
\end{center} 
\begin{center}
{\large Per K. Rekdal\footnote{Email address: perr@phys.ntnu.no.}$^{,a}$ 
and 
Bo-Sture K. Skagerstam\footnote{Email address: boskag@phys.ntnu.no.}$^{,a,b}$}
 \\[5mm] 
$^{a)}$Department of Physics, 
The Norwegian University of Science and Technology, \\
N-7491 Trondheim, Norway \\
$^{b)}$Theoretical Physics Division, CERN, CH-1211, Geneva 23, Switzerland
\end{center}
{\begin{center}
Physica Scripta (in press) \end{center}}

% 

%[-5mm]}} 
% 
% 
% 
\begin{abstract} 
\normalsize 
% 
%\vspace{-5mm} 
% 
\vspace{5mm} 
\noindent 
A simple model of a two-mode non-resonant parametric amplifier 
is studied with special regard to 
non-classical features such as revivals  and squeezing. 
The methods used apply for  
an arbitrary  pump parameter. Detailed analytical and explicit  expressions
are given when the coupling of the two modes has
an harmonic time-dependence. Despite its simplicity the model exhibits
a very broad range of intricate physical effects. We 
show that quantum revivals are possible for a 
broad continuous range of physical parameters in the case 
of initial Fock states. For coherent states we 
find that such revivals are  possible only 
for certain discrete rational number combinations of the ratio of frequency
detuning and pump parameters. Correlation
effects are shown to be very sensitive to the initial state of the system.
\end{abstract} 
\newpage

%%%%%%%%%%%%%%%%%%%%%%%%%%%%%%%%%   introduction.tex (start)  %%%%%%%%%%%%%%%%%%%%%%%%%%%%%%%%

\setcounter{page}{1} 
%%%%%%%%%%%%%%%%%%%%%%%%%%%%%%%%%%%%%%%%% 
\bc{ 
\section{Introduction\label{sec:introd}}} 
\ec 
%%%%%%%%%%%%%%%%%%%%%%%%%%%%%%%%%%%%%%%%%% 
% 
\vspace{5mm} 
% 

%\noindent
 
The quantum-mechanical analysis of coupled harmonic oscillators enters 
into many fields of physics. In the design of detectors for 
gravitational radiation the dynamics of an harmonic oscillator and a 
transducer is e.g.  of importance (see e.g. 
Ref.\cite{Bocko&Onofri&96}).  The coupling of free quantum fields to 
classical external fields can also be thought of as a system of 
harmonic oscillators with, in general, space- and time-dependent 
couplings (see e.g. Ref.\cite{Skagerstam&94}). In the field of quantum 
optics similar systems of coupled harmonic oscillator systems are of 
great interest as e.g. in the description of parametric down-conversion of 
photons (see
e.g. Refs.\cite{Louis&61,Gordon&63,Mollow&Glauber&67,Mollow&67,
Walls&Reid&84, 
Barnett&Knight&85, Barnett&Knight&87,Homes&Milburn&Walls&89,Shum&86,
Brief&96}).  

%\noindent 

Our primary goal here is to study a simple interaction between a 
LC-circuit and one mode of the second quantized radiation field all 
without damping effects included. The corresponding free Hamiltonians 
are used to define reference states as e.g. the notion of a photon. In 
order to calculate photon-number distributions at any time normal 
ordering procedures become an essential ingredient in our 
study. In particular we have, in terms of a canonical rescaling of 
variables, considered the following classical Hamiltonian

\be 
\label{eq:LC} 
H=\frac{1}{2}P_a^2 + \frac{1}{2}\omega_a^2 Q_a^2 + \frac{1}{2}P_b^2+  
\frac{1}{2}\omega_b^2Q_b^2 
-g\left(\sqrt{\frac{\omega_a}{\omega_b}}Q_aP_b + 
 \sqrt{\frac{\omega_b}{\omega_a}}P_aQ_b\right) ~~~, \ee 
where the canonical momentum $P_a$ is proportional to the electric current 
$I= dQ_a/dt$ of the LC-circuit with inductance $L$ and capacitance $C$ 
($\omega_a = 1/\sqrt{LC}$). $P_b$ and $Q_b$ describe the harmonic oscillator 
of the singe-mode of the radiation field with frequency $\omega_b$. The last 
interaction term is a simple form for the interaction of the radiation-field 
with the LC-circuit in terms of a real coupling constant g. The particular 
form of the interaction term in \eq{eq:LC} is motivated by the fact that 
a minor extension of this model can also be used to describe non-resonant 
parametric down-conversion, as will be discussed in more detail Section 
\ref{sec:dynsys}. \eq{eq:LC} is also the Hamiltonian for a two-dimensional 
electrically charged harmonic oscillator, in general non-isotropic, 
with an external 
magnetic field $B=g(\sqrt{\omega_b /\omega_a}- \sqrt{\omega_a /\omega_b} )$ 
perpendicular to the  ab-plane and with angular frequencies $\omega_a^2 
\rightarrow \omega_a^2 - g^2\omega_b /\omega_a$ and $\omega_b^2 \rightarrow 
\omega_b^2 - g^2\omega_a /\omega_b$. If the original angular frequencies 
$\omega_a$ and $\omega_b$ are equal the harmonic oscillator becomes isotropic 
and the magnetic field is zero.  
% 

%\noindent

 Despite the very simple structure of the Hamiltonian \eq{eq:LC} we will see 
that the corresponding quantum dynamics can lead to quite intriguing
physics. 
In view of the impressive experimental developments 
in quantum optics the system under consideration  may actually be realized in 
the laboratory. 

%\noindent

The paper is organized as follows. In Section 2 the
dynamical system is defined in more detail. 
An exact treatment of the time-evolution operator is presented in
Section 3. 
The results obtained turn out to have a broad range of applicability. 
Quantum revivals and their dependence of initial states are discussed
in Section 4. 
In Section 5 non-classical  features are studied in terms of
correlation functions. 
The effect of detuning on the squeezing properties of a parametric
amplifier
 is studied in Section 6. 
Signal-to-noise aspects are discussed i Section 7 and some final remarks 
are given in Section 8. 
\vspace{5mm} 

%%%%%%%%%%%%%%%%%%%%%%%%%%%%%%%%%   introduction.tex (end)  %%%%%%%%%%%%%%%%%%%%%%%%%%%%%%%%

%%%%%%%%%%%%%%%%%%%%%%%%%%%%%%%%%   dyn_sys.tex (start)  %%%%%%%%%%%%%%%%%%%%%%%%%%%%%%%%

%%%%%%%%%%%%%%%%%%%%%%%%%%%%%%%%%%%%%%%%%%%%%%%%%%%%%%%%%%%%%%%%%%%%%%%%%%%%%%%%%%%%%%%% 
% 
\bc 
{\section{The Dynamical System\label{sec:dynsys}}} 
\ec 
% 
%%%%%%%%%%%%%%%%%%%%%%%%%%%%%%%%%%%%%%%%%%%%%%%%%%%%%%%%%%%%%%%%%%%%%%%%%%%%%%%%%%%%%% 

%\noindent
 
The  coupled harmonic oscillators we consider are described by 
the two-mode Hamiltonian 
\be 
\label{H_tot} 
H=H_{a} +H_{b} +H_{int}~~, 
\ee 
where 
\bea 
H_a = \omega_a(n_a +\frac{1}{2}) ~~,\nonumber \\ 
H_b = \omega_b(n_b + \frac{1}{2}) ~~, 
\eea 
in natural units ($\hbar = c = 1$) and where $n_a \equiv a^\da a$ and $n_b \equiv b^\da b$. 
The interaction between the  $a$ and $b$ mode is chosen to be of the form 
 
\be \label{H_interaction} 
H_{int} = i\left[g(t)~ab - g^*(t)~a^{\da} b^{\da} \right] ~~, 
\ee 
where $g(t)$
    is a time-dependent coupling. If $g(t)=g$ we recover the
system described by \eq{eq:LC}.
The  interaction
Hamiltonian \eq{H_interaction} describes a two-mode non-degenerate parametric 
down-conversion process, where the modes are known as ``signal'' and ``idler'' and  
where the pump parameter $g(t)$ then 
describes an arbitrary classical pump field (for reviews see e.g.  
Refs.\cite{Shum&86,Yariv&67,Walls&Milburn&95,Mandel&Wolf&95,Scully&Zubairy&96}).  
% 
% 

%\noindent

 The Hamiltonian (\ref{H_interaction}) can be written in the form 
\be \label{H_int_time} 
H_{int} = i ~ \left[g(t) K_{-} - g^*(t) K_{+} \right]~~, 
\ee 
where 
\be 
K_{-} \equiv ab~~,~~K_{+}\equiv a^\da b^{\da}~~,~~K_{0}\equiv \frac{a^\da a + bb^{\da}}{2}~~~, 
\ee 
span the Lie algebra of the non-compact group $SU(1,1)$, i.e. 
\be\label{LSU(1,1)} 
[K_{+},K_{-}]=-2K_0~~,~~[K_{0},K_{\pm}]=\pm K_{\pm} ~~. 
\ee 
The Casimir operator of the Lie-algebra \eq{LSU(1,1)} is given by 
$C= K_0^2 - K_{+}K_{-} - K_0 = \Phi (\Phi +1)$, where we can choose 
$-2 \Phi = \mid n_a - n_b\mid + 1 $. We therefore see that we have a realization of 
the unitary representation $D^+(\Phi)$ of $SU(1,1)$ 
\cite{Bargmann&48etc}. 
% 
%We observe that if $\omega=0$ we recover the 
%Hamiltonian in \eq{eq:LC}. Hence, our Hamiltonian can be interpreted as a 
%model for the interaction between a LC-circuit and one mode of a second quantized 
%radiation field.  
% 
We observe that when we make the identification $a=b$, the Hamiltonian
describes a {\it degenerate} parametric amplifier. 
This corresponds to a realization of the $SU(1,1)$ Lie-algebra 
\eq{LSU(1,1)} with $K_{-}=a^{2}/2, K_{+}=(a^{\da})^{2}/2$ and $K_{0}= 
(a^\da a + aa^{\da})/4$ which also constitutes a realization of a discrete unitary 
representation of $SU(1,1)$ \cite{Dattoeta&86}. 
\vspace{5mm}

%%%%%%%%%%%%%%%%%%%%%%%%%%%%%%%%%   dyn_sys.tex (end)  %%%%%%%%%%%%%%%%%%%%%%%%%%%%%%%%

%%%%%%%%%%%%%%%%%%%%%%%%%%%%%%%  time_evol_op.tex (start)  %%%%%%%%%%%%%%%%%%%%%%%%%%%%%%%%

\vspace{5mm} 
 
%%%%%%%%%%%%%%%%%%%%%%%%%%%%%%%%%%%%%%%%%%%%%%%%%%%%%%%%%%%%%%%%%%%%%%%%%%%%%%%%% 
% 
\bc{ 
\section{ The Time-Evolution Operator \label{sec:solution}} 
}\ec 
% 
%%%%%%%%%%%%%%%%%%%%%%%%%%%%%%%%%%%%%%%%%%%%%%%%%%%%%%%%%%%%%%%%%%%%%%%%%%%%%%% 
\vspace{5mm}

%\noindent

 The Hamiltonian in the interaction picture and in terms of
the operators  
$K_{+}$ and $K_{-}$  reads 
\be \label{h_op} 
 H_{int}^I(t) = i\left[g(t) e^{-i(\omega_a + \omega_b)t} K_-  -  
                 g^*(t) e^{i(\omega_a + \omega_b)t} K_+  \right]~~~, 
\ee 
\noindent  
 with the corresponding Schr\"{o}dinger equation

\be \label{schro_eq} 
i \frac{d}{d t}U_I(t) = i\left[~{\tilde g}(t)   K_-  -  
                       {\tilde g}^*(t) K_+ ~  \right]~U_I(t) ~~, 
\ee 
 and where we have defined 
${\tilde g}(t) \equiv g(t) e^{-i(\omega_a + \omega_b)t}$.  
An exact analytical expression for the time-evolution operator $U_I(t)$  
can be obtained as follows. Due to the presence of the 
Lie-algebra of $SU(1,1)$ we  employ the Wei-Norman technique  
\cite{Wei&Norman&63,Dattoeta&86} and express the time-evolution
operator in the following  form 
\be \label{U_int} 
U_I(t)=e^{A_+(t) K_+} e^{2A_0(t) K_0} e^{A_-(t) K_-}~, 
\ee 
since this is a convenient choice when computing expectation values and 
probability amplitudes. If we substitute \eq{U_int} into the Schr\"{o}dinger 
equation \eq{schro_eq} we find that $A_+(t)$, $A_-(t)$ and $A_0(t)$
must satisfy the following 
system of non-linear differential equations: 
\bea \label{eqsystem} 
%\left 
\frac{d}{dt} A_+(t) &=& {\tilde g}(t)  A_+^2(t) - {\tilde g}^*(t)~, \nonumber \\ 
\frac{d}{dt} A_0(t) &=& {\tilde g}(t)  A_+(t)~,  \\ 
\frac{d}{dt} A_-(t) &=& {\tilde g}(t)  e^{2 A_0(t)}~, \nonumber 
%\right\} 
\eea 
with the initial conditions $A_+(0)=A_-(0)=A_0(0)=0$.   
These initial conditions ensure that $U_I(0)=1$. 
Eqs. (\ref{eqsystem}) are actually valid for the Hamiltonian 
Eq.(\ref{H_interaction})  
in general and  can e.g. be studied by making use of various numerical
methods. In the present paper we will, however, consider a simple
situation in which case analytical methods are at hand, i.e. we consider  a 
coupling $g(t)$ with harmonic time-dependence 
 
\be \label{g_here} 
 g(t)= g e^{i\omega t}~~, 
\ee 
where  $g$ is a real constant. For parametric  
down-conversion processes the frequency  $\omega$ is then referred to as the  
pump frequency. In the course of preparation of the present paper we have observed that
in the special case of the time-dependence of \eq{g_here} algebraic
methods can be used ¨
to solve for
\eq{schro_eq} (see e.g. Ref.\cite{Zahler&Aryeh&91}). 
To the extent one can compare our methods and results 
with those of Ref.\cite{Zahler&Aryeh&91} we find agreement. 
We, however, once more stress that our methods
in principle can be used for an arbitrary time-dependence of $g(t)$. 
It is convenient to define 
\be 
 \Omega = \omega - \omega_a - \omega_b ~~~. 
\ee 
When the frequencies $\omega_a$ and  $\omega_b$ sum up to the classical pump 
frequency $\omega$, i.e. $\Omega=0$, we have parametric resonance. In this 
case the Schr\"{o}dinger equation is easy to solve (see e.g. Ref.  
\cite{Walls&Milburn&95}). 

Let us now consider the more general situation $\Omega \neq 0$, in which case  
the interaction Hamiltonian constitutes  
a simple model of a detuned "broad-band" pump in 
parametric amplification. Within first-order perturbation theory, in
which case the 
time-ordering procedure of the time-evolution operator is not important, 
the broad-band pump  down-conversion has  been considered in great detail in the 
literature
\cite{Klyshko&69,Ghosh&etal&86,Ou&etal&89,Teich&etal&94,
Teich&etal&96,Rubin&etal&94,Milonni&etal&96,Casado&etal&97,Grice&Walmsley&97}.
Here we are 
interested in an exact 
treatment of the time-evolution operator, i.e. we have to solve 
Eqs. (\ref{eqsystem}) with the harmonic time-dependence \eq{g_here}. 
An exact treatment can, e.g., be used to investigate unitarity effects
(i.e. including not only the vacuum and the two-photon states 
but all possible intermediate states) of photon correlation experiments.

%\noindent
 
A important  point here is that we have to solve Eqs. (\ref{eqsystem}) 
separately for three different cases:  
$k^2 < 1$, $k^2> 1$ and $k^2=1$, where we have defined the dimensionless   
parameter

\be 
k \equiv \frac{\Omega}{2 g}~~. 
\ee 
In the case $k^2 < 1$, the solution has the form 
\bea
\label{firstsoul} 
A_+(t) &=& -~ e^{-i \Omega t} \left [ \sqrt{1-k^2 } \tanh \left ( 
           gt \sqrt{1-k^2} - i\gamma \right ) + i k 
  \right ]~, \nonumber \\ 
A_-(t) &=&  \sqrt{1-k^2} \tanh \left ( 
           gt \sqrt{1-k^2 } - i\gamma \right ) + i k ~, 
           \\ 
A_0(t) &=& - \ln \left [ \frac{ \cosh \left ( gt \sqrt{1-k^2 } 
             - i \gamma \right )}{\cos \gamma}  \right ] -
       \frac{i\Omega}{2}t ~,\nonumber 
\eea 
where 
\be \label{tan_alpha} 
\tan\gamma = \frac{ k }{\sqrt{1 - k^2  }} ~. 
\ee 
 The solution in the case $k^2 > 1$ is of the same form as in the case
 above with the 
 substitutions $ \gamma \rightarrow -\gamma (\equiv \delta) $, 
 $\tanh \rightarrow \cot $ and $\cos \rightarrow i \sinh $. The explicit 
 solution of the Eqs. (\ref{eqsystem}) is then  
\bea  \label{A_bigger_1} 
A_+(t) &=& -~  e^{-i \Omega t} \left [ \sqrt{ k^2 -1 } \cot \left ( 
               gt \sqrt{ k^2 -1} + i\delta \right ) + ik \right ]~, \nonumber \\ 
A_-(t) &=&    \sqrt{ k^2 -1 } \cot \left ( 
               gt \sqrt{k^2-1} + i\delta \right ) + ik ~,     \\ 
A_0(t) &=&  - \ln \left [ \frac{ \sin \left ( gt \sqrt{ k^2 -1} 
            +  i \delta \right )}{\sin ( i \delta )}  \right ] - \frac{i\Omega}{2}t 
             ~,\nonumber 
\eea 
where now 
\be  \label{coth_beta}  
\coth\delta = \frac{ k  }{\sqrt{ k^2 -1}} ~. 
\ee 
If $k^2 = 1$ it is particularly easy to show that 
\bea 
\label{lastsoul}
A_+(t) &=& -~  e^{-i \Omega t} \left [ \frac{1}{gt + i} + i \right ]~, \nonumber \\ 
A_-(t) &=&   \left [ \frac{1}{gt + i} + i \right ]  ~,\\ 
A_0(t) &=& - \ln \left [ 1 - i gt \right ] - \frac{i\Omega}{2}t ~,\nonumber 
\eea 
is the solution of the Eqs. (\ref{eqsystem}) with the correct initial 
conditions. We notice  that $A_+(t) = -~ e^{-i \Omega t} A_-(t)$ for
all three cases. 
 
%\noindent

 Due to the constraint of unitarity, i.e. $U_I^\da(t) =
U_I^{-1}(t)$, the functions 
$A_+(t)$, $A_-(t)$ and $A_0(t)$ are not independent. By making use of operator 
reordering techniques (for a pedagogical account see
e.g. Ref.\cite{Mufti&93}), the unitarity of $U_I(t)$ 
leads to the general conditions 
\bea \label{unitarity_sys} 
A_+^*(t) &=& - \frac{ A_-(t)}{e^{2 A_0(t)} -  A_-(t)A_+(t)}~, \nonumber \\ 
A_-^*(t) &=& - \frac{ A_+(t)}{e^{2 A_0(t)} -  A_-(t)A_+(t)}~, \\ 
 e^{-A_0(t)-A_0^*(t)} &=& 1 -  A_-(t)A_+(t) e^{-2A_0(t)}~~, \nonumber 
\eea 
and hence 
\be 
e^{A_0(t)-A_0^*(t)}= - \frac{A_-(t)}{A_{+}^{*}(t)} =  
- \frac{A_+(t)}{A_{-}^{*}(t)} ~~. 
\ee 
Using (\ref{unitarity_sys}) the following useful relationship follows 
\be \label{spec_eq} 
e^{-A_0(t) - A_0^*(t)}  \left [ 1 - |A_-(t)|^2 \right  ] = 1~~. 
\ee 
It is not  obvious that the explicit solutions \eq{firstsoul}
- \eq{lastsoul} for
 $A_+(t)$, $A_-(t)$ and $A_0(t)$ satisfy the Eqs. 
(\ref{unitarity_sys}). However, by using the fact that $A_+(t) = -~ e^{-i \Omega t} 
A_-(t)$ and various trigonometric relations, one can explicitly show that 
Eqs. (\ref{unitarity_sys}) are  fulfilled for all three cases $k^2 <
 1$, $k^2> 1$ and $k^2=1$. 
 
%\noindent 

With the solutions above it is now straightforward to calculate
various expectation values 
and transition amplitudes. In this context it is useful to obtain the annihilation 
operator $a_H(t)\equiv a(t)$ in the Heisenberg picture ($a(t=0)\equiv a$), i.e. 
\be 
a(t) = U(t)^{\da} a U(t)~, 
\ee 
where 
\be \label{U_tot} 
U(t) = U_0(t)U_I(t)~, 
\ee 
and 
\be \label{U_0} 
U_0(t)= \exp[-i(\omega_aa^{\da}a + \omega_b b^{\da}b + E_0/2 )t~]~~. 
\ee 
Here $E_0= \omega_a + \omega_b $ and  $U_I(t)$ is given by \eq{U_int}. 
Straightforward calculations now  gives 
\be \label{a_H_eq} 
a(t)= e^{-A_0^* - i \omega_a t}\left [a ~ - b^{\da}~ A_-^*(t) \right ]~, 
\ee 
from which $b(t)$ follows by the substitution $a\leftrightarrow b$. The creation 
operator $a^{\da}$ follows from \eq{a_H_eq} by taking the Hermitian 
conjugate. 
When $k^2=0$ this solution for the operator $a(t)$ reduces to the wellknown result  
(see e.g. Ref.\cite{Walls&Milburn&95}) 
\be 
  a(t) = ~  a\cosh(gt) -  b^{\da}\sinh(gt) ~~. 
\ee 
\noindent The commutator $ [ a(t), a^{\da}(t)]$ 
is in general given by 
\be \label{kom_tor} 
\left [ a(t), a^{\da}(t) \right ] =  e^{-A_0(t) - A_0^*(t)} 
                                 \left [ 1 - |A_-(t)|^2 \right  ]~~. 
\ee 
Due to \eq{spec_eq} the quantum condition $\left [ a(t), a^{\da}(t) 
\right ] = 1$ is satisfied for all times, as it should be.  

The Hamiltonian of the system is such that  
$a^{\da}(t)a(t) -b^{\da}(t)b(t)$ is a conserved quantity. We obtain 
\bea 
\label{n_a_operator}
    && ~~~~~~~~~~~~~~~~~~~~~~n_{a}(t) \equiv  a^{\da}(t)a(t) = \nonumber \\
 && a^{\da}a - A_-(t)K_- + A_+(t)e^{-2A_0(t)}
\left(K_+ -2A_-(t)K_0 + A_{-}^{2}(t)K_- \right)~~. 
\eea 
%\da her istedenfor * 
%
Since $n_{b}(t) \equiv b^{\da}(t)b(t)$ can be obtained from 
\eq{n_a_operator} by means of the substitution 
$a\leftrightarrow b$ it follows  
immediately that $n_{a}(t)- n_{b}(t)= n_{a}(0) - n_{b}(0)$. 
The time-dependence of other  
combinations of annihilation and creation operators can be obtained 
in s similar manner. 
\vspace{5mm} 
% 
% 
%\end{document} 
 
\vspace{1cm} 
%%%%%%%%%%%%%%%%%%%%%%%%%%%%%%%%%%%%%%%%% 
\bc{ 
\section{Quantum Revivals} 
}\ec 
%%%%%%%%%%%%%%%%%%%%%%%%%%%%%%%%%%%%%%%%% 
%\noindent
 
In this section we focus our attention on quantum revivals, i.e. when 
the initial state is reproduced by the time-evolution (for a reviews
see e.g. Refs.\cite{Averbukh&etc&91}). 
The time-evolution operator \eq{U_int} is written in a form 
that makes the evaluation of transition amplitudes   
$\langle f|~U(t)~ |i\rangle$ straightforward. It is clear from the 
Eqs.(\ref{A_bigger_1}) that exact quantum revivals will occur for Fock states
only if $k^2 >1 
$ and at revival times $t_{rev}$ such that
\be \label{eq_period} 
 g t_{rev} \sqrt{k^2-1} = n \pi ~~, 
\ee 
where $n$ is a positive integer. This is so since the corresponding
transition probability is  independent of $\Omega $.
 The phase in the transition 
amplitudes due to the free time-evolution will not effect the transition
probabilities, at least not for initial and final Fock-states. For   
coherent states we will, however,  find it convenient to evaluate the 
transition 
probabilities in a basis in which case the free time-evolution is absent. 
 For coherent
states  it then turns out that \eq{eq_period} is extended in such a
way  that only certain
rational numbers of $k^2$ will lead to revivals. Here we therefore see
a remarkable difference between initial and final  Fock states and
coherent states respectively with regard to quantum revivals. We now
treat these special cases in more detail.
\vspace{5mm} 
%%%%%%%%%%%%%%%%%%%%%%%%%%%%%%%%%%%%%%%%% 
\bc{ 
\subsection{Fock States} 
}\ec 
%%%%%%%%%%%%%%%%%%%%%%%%%%%%%%%%%%%%%%%%% 

%\noindent
 
The two-mode initial state has now the form $|i \rangle = |r \rangle_a \otimes 
| s \rangle _b \equiv |r,s \rangle$. We also use the notation 
$\langle f| = {_{b}\langle} m| \otimes {_{a}\langle }n| \equiv \langle 
m,n| $. The transition amplitudes to be derived are $c_{mn}(t;r,s) \equiv 
\langle m,n| U_I(t)|r,s \rangle$. 
The probabilities to be calculated are then $p_{mn}(t;r,s) \equiv | 
c_{mn}(t;r,s) |^2$. We will suppress the initial state indices $r$ and
$s$ 
if it is clear from the context what initial state we are 
considering. The corresponding amplitudes $c_{mn}(t)$ can be found in closed 
form by making use of the results from Section \ref{sec:solution}, i.e. 
 \bea 
  c_{mn}(t) &=& \delta_{m,s-r+n}~\sqrt{r!s!m!n!} ~ e^{A_0(t)(s+r+1)} \nonumber \\ 
                & & \times \sum_{k=0}^{\mbox{\footnotesize  min}(r,s)} 
                    \frac{e^{-2kA_0(t)}}{(r-k)!(s-k)!}\frac{[A_-(t)]^k}{k!} 
                    \frac{[A_+(t)]^{n+k-r}}{(n+k-r)!}~~. 
\eea 
It follows from the actual expressions for  $A_0(t), A_+(t)$ and
$A_-(t)$ that $|c_{mn}(t)|^2$ does not depend on $\Omega $ and therefore
one obtain the revival times \eq{eq_period}.
% 
% 
%\noindent
  
We now present more explicit results in a few simple cases. If the  
initial state is the vacuum state, i.e. $r=s=0$, it  follows that 
\be 
\label{p_nn_00} 
        p_{mn}(t) = \delta_{mn}~e^{A_0(t)+A_0^*(t)} (|A_{+}(t)|^2)^n ~~. 
\ee 
It can now be shown explicitly that this probability distribution is 
properly normalized, i.e. $\sum _{m,n=0}^{\infty}p_{mn}(t) = 1 $ for all times.
$p_{mn}(t)$ can be explicitly evaluated by making use of trigonometric identities.
One can then derive the useful results 
\be \label{x_label}                                                           
  x(t) \equiv e^{-A_0(t)-A_0^*(t)} = 
\frac{\sinh^2(gt\sqrt{1-k^2})+1-k^2}{1-k^2}~~, 
\ee 
\be \label{y_label} 
  y(t)\equiv |A_{+}(t)|^2 =|A_{-}(t)|^2 = 
   \frac{\sinh^2(gt\sqrt{1-k^2})}{\sinh^2(gt\sqrt{1-k^2})+ 1-k^2}~~,         
\ee 
and 
\be \label{z_label} 
  n_0(t)\equiv -A_-(t) A_+(t) e^{-2 A_0(t)} = \frac{\sinh^2(gt\sqrt{1-k^2})} 
  {1-k^2}~~, 
\ee 
valid for all $k^2$.
The reason why we use the notation $n_0(t)$ is 
the fact that \eq{z_label} is 
nothing else but the expectation value of the number operator $n_{a}(t)$,
as given by
\eq{n_a_operator} (or $n_{b}(t)$),  in the initial vacuum state.
We notice that the last three equations are connected in the following way:  
$x(t) y(t) = n_0(t)$ and $x(t) = 1 + n_0(t)$.  

%\noindent

 The reduced density
matrix for the $a$-modes (or the $b$-modes) is again given in terms of
\eq{p_nn_00}. As has been noticed before 
(see
e.g. Refs.\cite{Mollow&Glauber&67,Walls&Reid&84,
Barnett&Knight&85,Barnett&Knight&87,
Homes&Milburn&Walls&89}) 
the reduced density matrices describe a system at finite temperature
$T=1/ \beta $ such that 
$\exp(-\beta\omega) = y(t)$,
where $\omega = \omega_{a}$ (or $\omega =\omega_{b}$ for the
$b$-modes), i.e. if $k^2 < 1$ the
temperature approaches infinity in the limit of large $gt$.
This interpretation has, of course, a limited 
range of validity.
 
%\noindent
  
\begin{figure}[t] 
\unitlength=1mm 
\begin{picture}(160,80)(0,0) 
\includegraphics{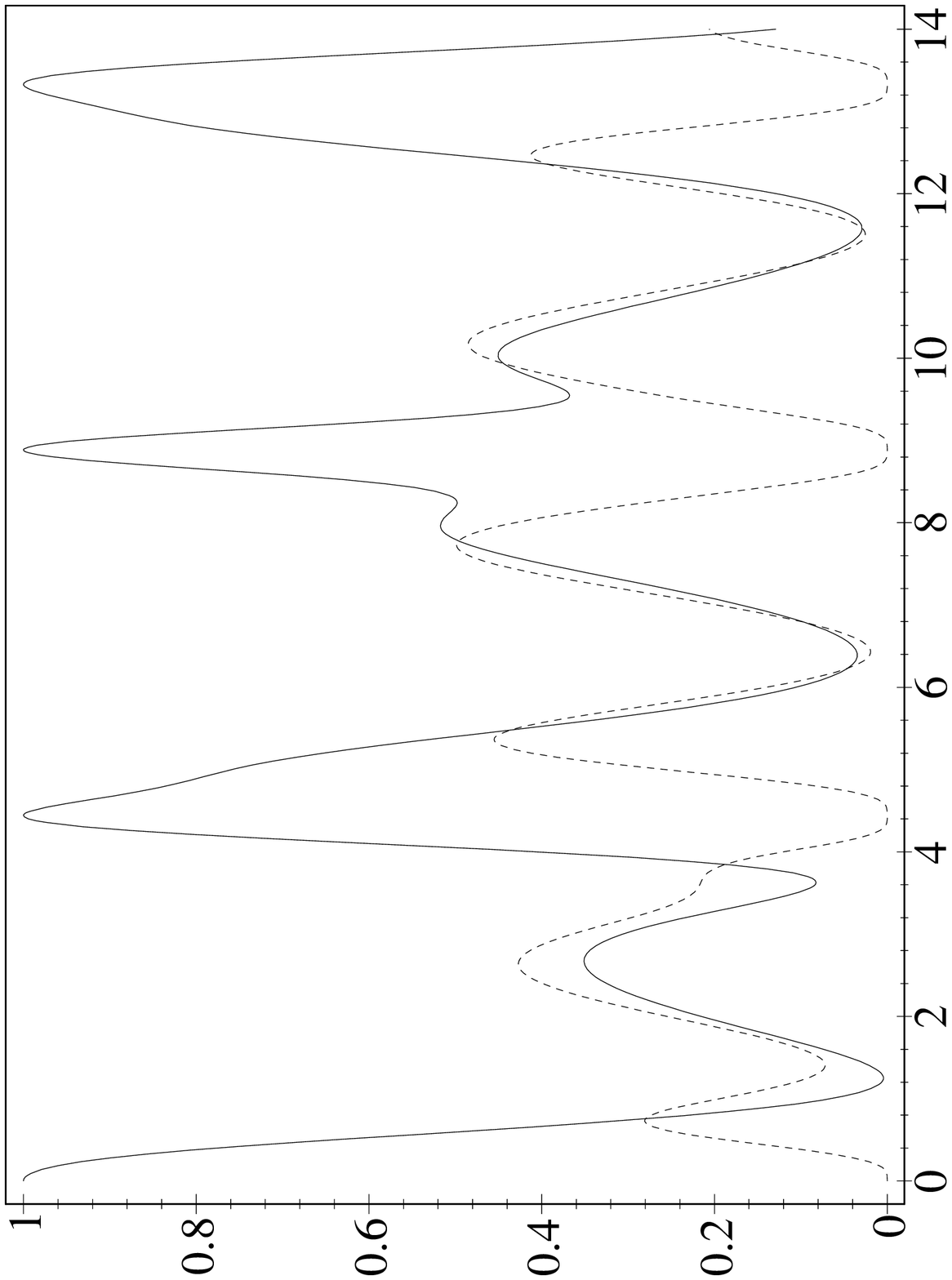} 
   \put(79,-5){\normalsize  \boldmath$gt$} 
   \put(0,50){\normalsize   \boldmath$p_{mn}(t)$} 
\end{picture} 
\vspace{2mm} 
\figcap{ The probabilities $p_{11}(t;1,1)$ (solid curve) and 
         $p_{33}(t;1,1)$ (dotted curve) when $k^2=1.5$ and with an
         initial vacuum state. The 
         corresponding revival-times are integer multiples of  
         $\pi\sqrt{2}\approx 4.44$. 
\label{fig_p_nn_11_1_5} } 
\end{figure}

\begin{figure}[t] 
\unitlength=1mm 
\begin{picture}(160,80)(0,0) 
\includegraphics{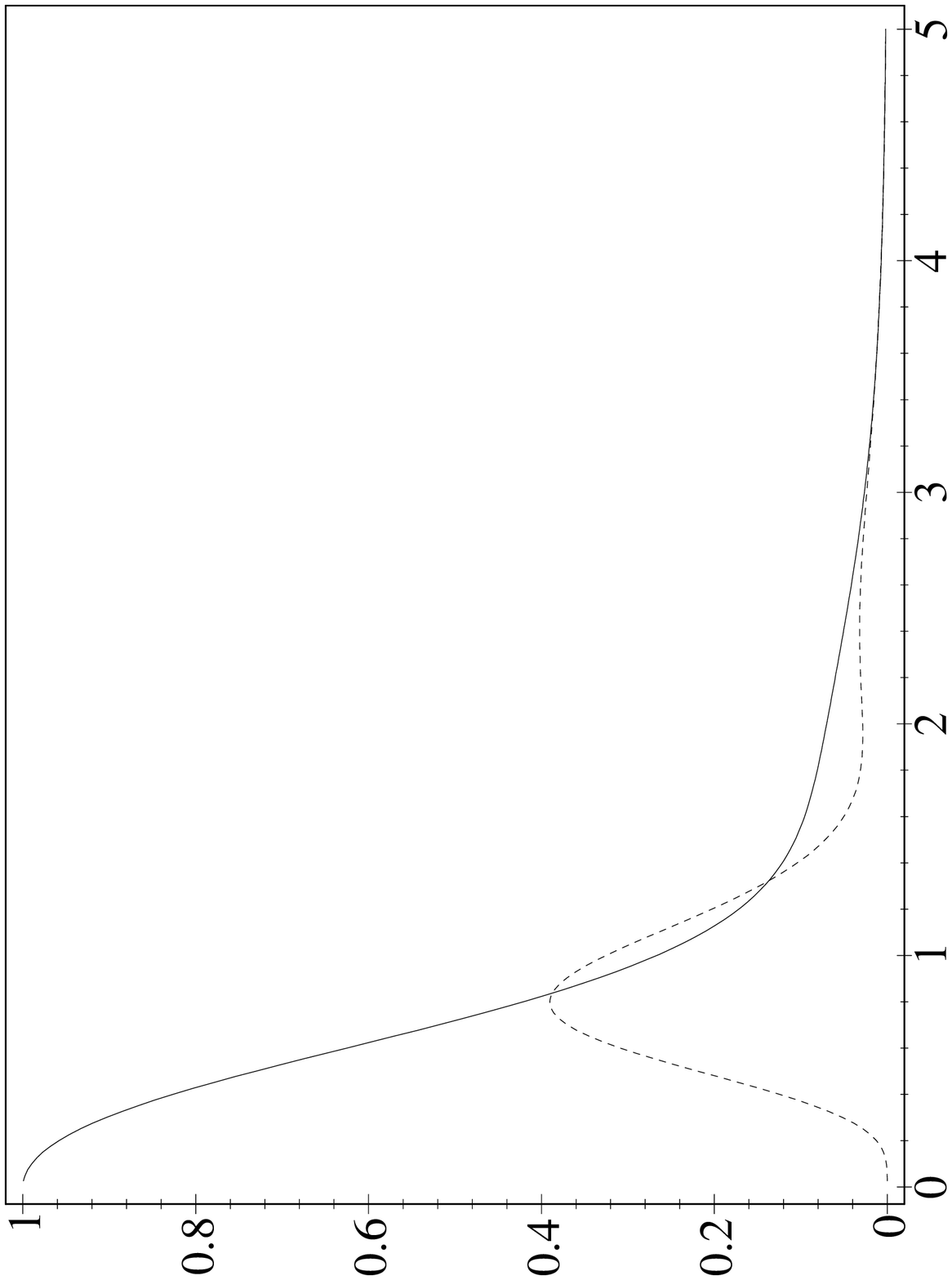} 
   \put(79,-5){\normalsize  \boldmath$gt$} 
   \put(0,50){\normalsize \boldmath$p_{mn}(t)$} 
\end{picture} 
\vspace{2mm} 
\figcap{The probabilities $p_{11}(t;1,1)$ (solid curve) and 
        $p_{33}(t;1,1)$ (dotted curve) when $k^2=0.5$ and with an
        initial vacuum state.
\label{fig_p_nn_11_0_5} 
} 
\end{figure}

With $|i \rangle = |1,1\rangle$, the corresponding  
probabilities are given by 
\be 
\label{p_nn_11} 
  p_{mn}(t) = \delta_{mn} \left ( \frac{n^2}{x^2(t)}y^{n-1}(t)-
    \frac{2n}{x(t)}y^{n-1}(t) 
    + y^{n+1}(t)
   \right )\frac{1}{x(t)}~~~, 
\ee 
and again it follows that $p_{mn}(t)$ is properly normalized. 
Furthermore we have $\langle i|n_a(t)|i\rangle = 1+2n_0(t)$.
 The probabilities in \eq{p_nn_11} for $n=m=1$ and $n=m=3$ are shown in 
\fig{fig_p_nn_11_1_5} for 
$k^2=1.5$ in which case quantum revivals are predicted. 
In \fig{fig_p_nn_11_0_5} the same  
probabilities are shown, but now for the case $k^2=0.5$. 
These figures reflect the typical physical situation: 
For $k^2<1$ the probability has initially a peak structure 
and then it decays exponentially with time. Actually all probabilities
$p_{mn}(t)$ will then tend to zero for sufficiently large times in such a way
that  their sum still is one, i.e. $\sum _{m,n=0}^{\infty}p_{mn}(t) = 1$. 
  On the other hand, if $k^2>1$ the probability exhibit a totally 
different oscillatory behavior  
for all times with quantum revivals as predicted by
\eq{eq_period}. 
 The reduced density matrices can be computed as above.
For large $gt$ and $k^2 < 1$ one finds that the reduced density matrices
correspond to the same thermal distribution  as  for the initial vacuum state as 
discussed above.

% 
% 
% 
%\vspace{2cm} 
% 
% 
% 

%\vspace{2cm} 
% 
% 
 
%\noindent
 
\begin{figure}[t] 
\unitlength=1mm 
\begin{picture}(160,80)(0,0) 
\includegraphics{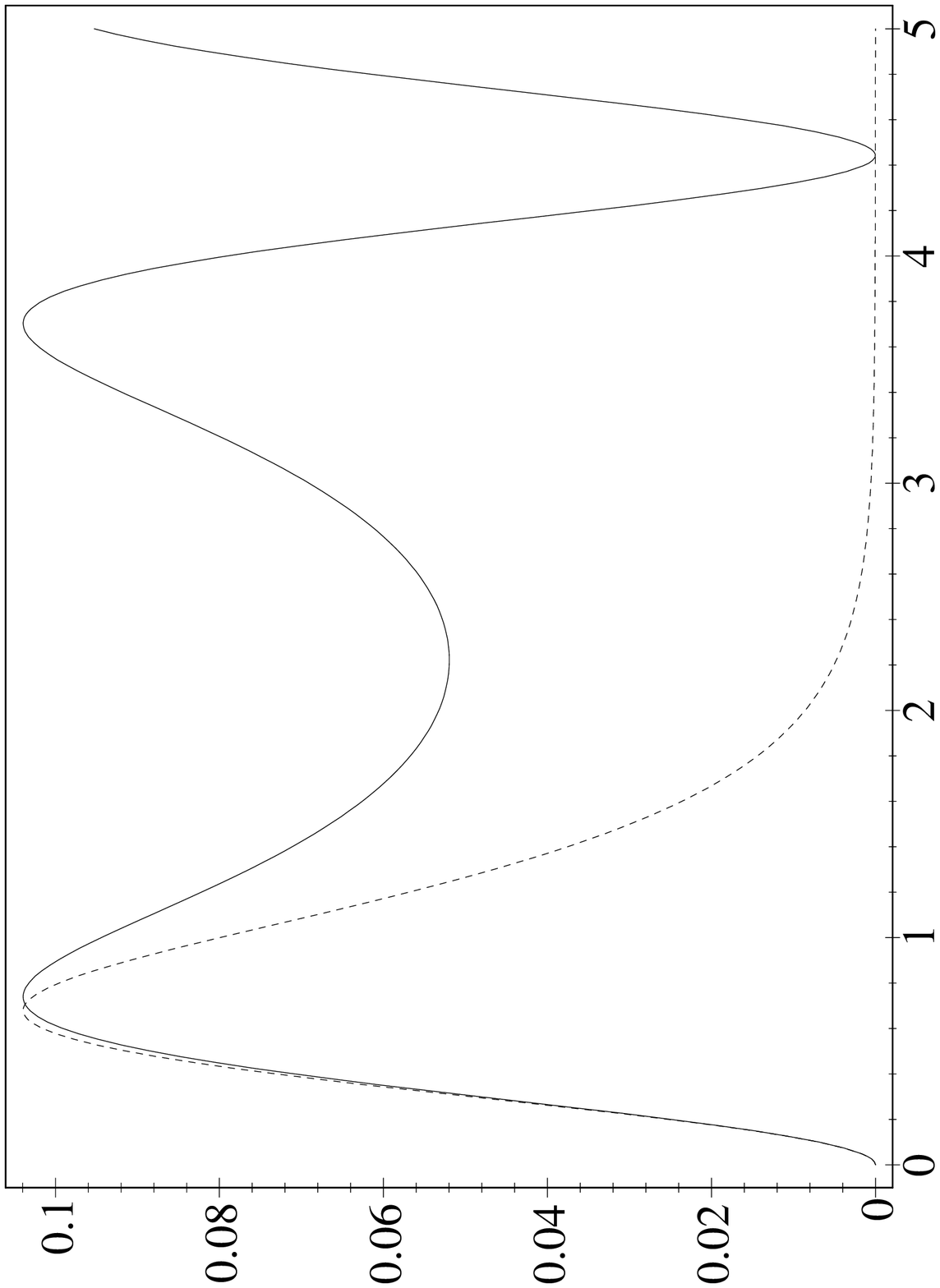} 
   \put(75,-5){\normalsize  \boldmath$gt$} 
   \put(0,48){\normalsize \boldmath$p_{mn}(t)$} 
\end{picture} 
\vspace{2mm} 
\figcap{ The probability $p_{mn}(t)$ in \eq{p_mn_0Psi} 
                  when $n=2$ and $m=1$ for the two cases 
                  $k^2=1.5$ (solid curve) and $k^2=0.5$ (dotted curve) 
                  and when the $a$-photons initially are Poisson distributed 
                  ${\cal P}_n = |\alpha|^{2n} e^{-|\alpha|^2}/n!$ with 
                  amplitude $\alpha = 0.85$.
\label{fig_p_mn_0Psi} } 
\end{figure}

Let us now assume that there is initially no $b$-quanta, that 
is  $|i \rangle = |\psi \rangle_a \otimes | 0 \rangle _b $,
 where the initial $a$-mode state is a general pure state described by 
\be \label{psi_BF} 
|\psi \rangle _a 
  = \sum_{s=0}^{\infty} \sqrt{{\cal P}_s}~ e^{i \phi_s} |s \rangle_a~~. 
\ee 
Here ${\cal P}_s$ is the probability that the $a$-mode is in the Fock state 
$|s \rangle _a$ and $\phi_s$ is the corresponding phase. As above, 
the final states $U_0 |r,s \rangle$ and $|r,s \rangle$ give the same 
result for the probability.   
The probability $ p_{mn}(t) \equiv |\langle m,n|  ~ U_I(t) ~ |\psi,0  
\rangle|^2$ is given by 
\be \label{p_mn_0Psi} 
 p_{mn}(t) = 
{\cal P}_{n-m} \frac{n!}{(n-m)!~m!} 
\frac{y^m(t)}{x^{(n-m + 1)}(t)}~~, 
\ee 
independent of the phase  $\phi _n$. 
Again one can analytically verify that  $\sum_{m,n=0}^{\infty}
p_{mn}(t) = 1$ 
for all times when $\sum_{n=0}^{\infty} {\cal P}_n  =1$ for 
all three cases $k^2> 1$, 
$k^2< 1$ and  $k^2 = 1$. The expectation values of the
number operators for the $a$  and $b$-modes are 
 $\langle i|n_a(t)|i\rangle = \langle n_a \rangle +n_0(t)[\langle n_a
 \rangle+1]$ 
and  
$\langle i|n_b(t)|i\rangle = n_0(t)[\langle n_a \rangle+1]$
respectively. Here we have that
$\langle n_a \rangle = \sum_{n=0}^{\infty}n{\cal P}_n $.
If in particular the $a$-mode is Poisson distributed 
the probability $p_{12}(t)$ is 
shown in \fig{fig_p_mn_0Psi} for $k^2 = 0.5$ and $k^2 = 1.5$.
When $k^2>1$ the probability oscillates between $0$ and the maximum
value of  
$p_{12}(t)$, i.e. $8|\alpha|^2\exp(-|\alpha|^2)/27 $ which is close to
$0.104$ for the parameters corresponding to \fig{fig_p_mn_0Psi}. 
Furthermore, the case $k^2=1$ is very much like case $k^2<1$.

 The reduced density matrices for the $a$- and  $b$-modes are again
diagonal with the matrix elements
\be \label{reduced_a}
p_m(b;t) = \sum_{n=0}^{\infty}p_{mn}(t)
= \sum_{l=0}^{\infty}{\cal P}_l 
\frac{(l+m)!}{m!~l!}\frac{y^m(t)}{x^{l+1}(t)}~~~,
\ee
for the $b$-modes and
\be \label{reduced_b}
p_n(a;t) = \sum_{m=0}^{\infty}p_{mn}(t)
= \frac{1}{x^{n+1}(t)} \sum_{m=0}^{n}{\cal P}_{n-m}
\left(\begin{array}{c}n\\m\end{array}\right)
n^m_0(t)~~~.
\ee
for the $a$-modes. For large $gt$ and if $k^2 <1$ one finds again that both
modes are thermally distributed with the same distribution as in the
case of an initial vacuum state. 
%\vspace{2cm} 

\vspace{5mm}  
% \newpage
%%%%%%%%%%%%%%%%%%%%%%%%%%%%%%%%%%%%%%%%% 
\bc{ 
\subsection{Coherent State} 
}\ec 
%%%%%%%%%%%%%%%%%%%%%%%%%%%%%%%%%%%%%%%%% 
%\noindent
 
\begin{figure}[t]
\unitlength=1mm
\begin{picture}(160,80)(0,0)
\includegraphics{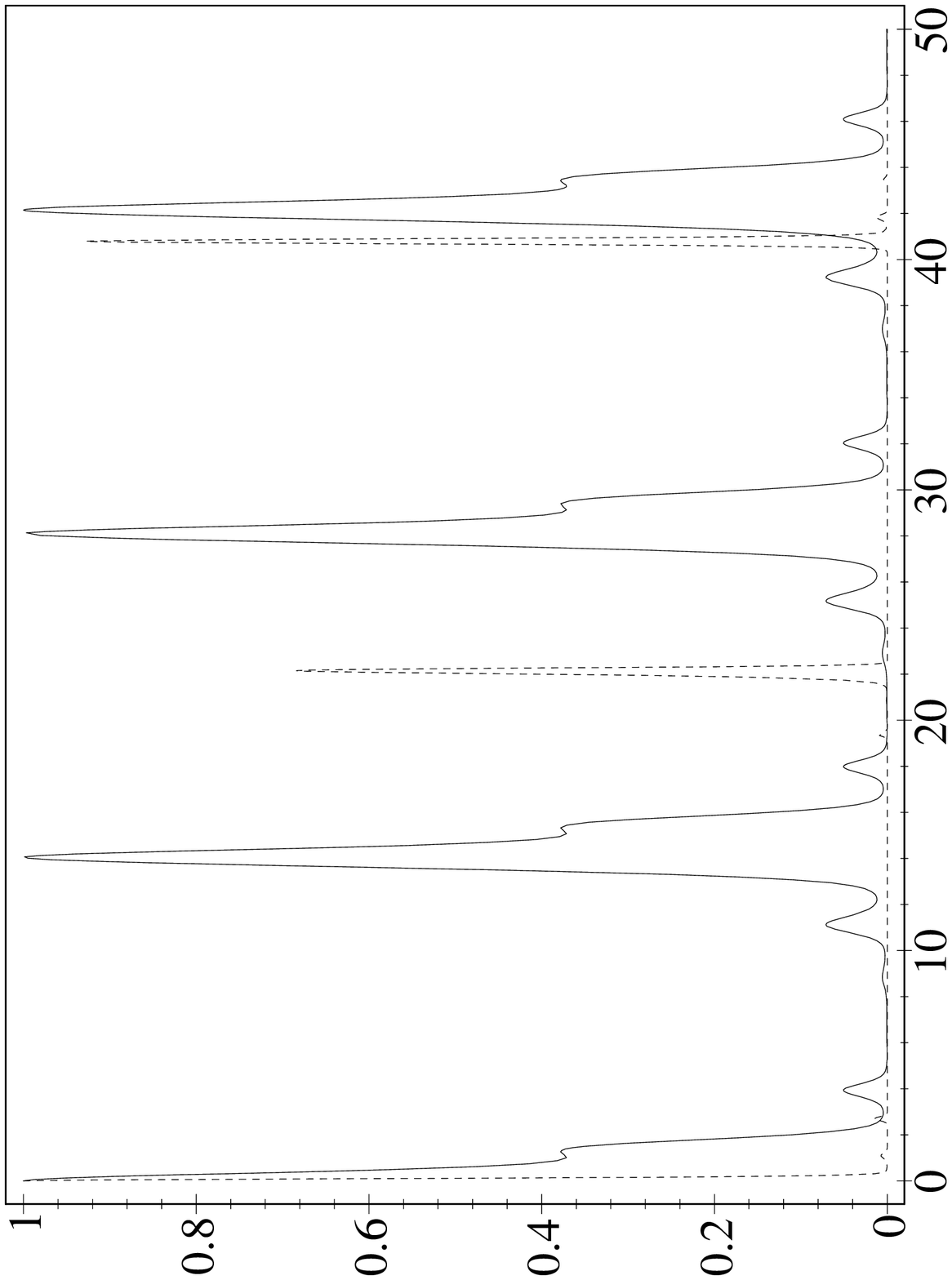}
   \put(79,-5){\normalsize  \boldmath$gt$}
   \put(0,50){\normalsize   \boldmath$p_{\beta \alpha}(t)$}
\end{picture}

\vspace{2mm}
\figcap{ The solid curve shows the probability $p_{\beta \alpha}(t)$
         when $k^2=9/5$ and  $\alpha=\beta=1$. The dotted curve 
         shows $p_{\beta \alpha}(t)$ when $k^2=\pi$ and  $\alpha=\beta=5$. 
         The revival times corresponding to the solid 
         curve are integer multiples of $gt_{rev}= \pi \sqrt{20} \approx 14.05$. 
\label{fig_p_zw} }
\end{figure}

We now consider initial and final coherent states.
The coherent state transition amplitudes are somewhat general in the 
sense that all Fock-state amplitudes can be obtained from them in a 
standard manner (see e.g. Ref.\cite{Klauder&Skagerstam&85}).
If the final state is a coherent state, then the presence of $U_0(t)$
is important since
$U_0(t)~ |\alpha \rangle _a \otimes |\beta \rangle_b = e^{-i E_0 t /2}~|\alpha 
    e^{-i \omega_a t}  \rangle _a \otimes |\beta e^{-i \omega_b t} \rangle_b$. 
    To exhibit quantum revivals we therefore find it convenient to use 
    $U_0(t)~|\alpha \rangle_a \otimes |\beta \rangle_b$ as a final state.    
    Let us, in particular,
 consider the case when the system is initially in the coherent state 
    $|i \rangle = |\alpha \rangle_a \otimes |\beta \rangle_b \equiv | 
    \alpha,\beta \rangle $ and final state $\langle f| = {_b\langle z|} 
    \otimes {_a\langle w|}~U_0^{\da}(t) \equiv \langle z,w|~U_0^{\da}(t)$. 
    The transition probability $p_{z w}(t) = |\langle z,w| U_I(t) 
    |\alpha,\beta\rangle|^2$ is then given by
\newpage
    \bea \label{p_ba_zw}
    p_{z w}(t) &=&  e^{\displaystyle -\left( |z|^2 + |w|^2 +
    |\alpha|^2 + |\beta|^2  \right )}
                e^{\displaystyle A_0(t) +  A_0^{*}(t)}
                 \\  && \nonumber
                 \times ~  
                 |e^{\displaystyle A_+(t)w^*z^*}|^{\displaystyle 2}
                 |e^{\displaystyle A_-(t)\alpha \beta}|^{\displaystyle 2}
                 |e^{\displaystyle w^* \alpha e^{A_0(t)}}|^{\displaystyle 2}
                 |e^{\displaystyle z^* \beta e^{A_0(t)}}|^{\displaystyle 2}~~~.
\eea
\eq{p_ba_zw} can be used as a 
generating function for the Fock state transition probabilities.
The  expectation value of the
number operator for the $a$-mode is
\be 
 \langle i|n_a(t)|i\rangle = \langle n_a \rangle +n_0(t)[\langle n_a
 \rangle +  \langle n_b
 \rangle   +1] - 2\mbox{Re}[\alpha\beta A_-(t)](1+n_0(t))~~~, 
\ee
and  for the $b$-modes
$\langle i|n_b(t)|i\rangle =   \langle i|n_a(t)|i\rangle +
 \langle n_b \rangle -\langle n_a \rangle $. Here we
have $\langle n_a \rangle = |\alpha|^2$ and 
$\langle n_b \rangle = |\beta |^2$.
In  contrast to the Fock case, this transition  probability  
   depends on the detuning frequency $\Omega$. To obtain exact quantum 
   revivals, the parameter $k^2$ can therefore not take an arbitrary value.
   To investigate the revival period of \eq{p_ba_zw} it is convenient to 
   rewrite the revival probability in the following form 
%
%\newpage
%
%
\bea \label{p_rewrite}
   p_{\beta \alpha}(t)& = & e^{\displaystyle 
                           -( |\alpha|^2 + |\beta|^2 )[ 2 - 
                        e^{ A_0(t)} - 
                        e^{ A_0^*(t)}]} \nonumber \\ &&
                        \times ~ | e^{\displaystyle \alpha  
                        \beta [A_-(t) + A_+^*(t)]}|^{\displaystyle 2}
        e^{\displaystyle A_0(t)+  A_0^*(t)}~~~.
\eea 
   By inspection of the actual expressions for $A_+(t)$ and $A_-(t)$ in 
   \eq{A_bigger_1} we see that the second factor in \eq{p_rewrite}
   is periodic when 

\be \label{param_k2}
   k^2 = \frac{1}{1 - (p/n)^2}~,
\ee
  where $n$ and $p$ are integers satisfying $n>p$, i.e. $k^2$ has to be a 
  rational number in order to obtain exact quantum revivals.
  From the first factor \eq{p_rewrite} we also see that 
  $A_0(t)$ has to be periodic in order to observe revivals.  
  Thus, we have further restrictions on the integers $n$ and $p$: if $n$ is 
  even, then $p$ must also be even. If $n$ is odd, then $p$ must also be odd.   
  The revival time corresponding to \eq{param_k2} is then
\be \label{coh_rev_time}
  g t_{rev} = \pi ~\sqrt{n^2 - p^2}~.  
\ee
   The solid curve in \fig{fig_p_zw} shows the probability $p_{\beta \alpha}(t)$ 
   when $k^2=9/5$, i.e. when $k^2$ satisfies \eq{param_k2} ($\alpha =
   \beta = 1$). The revival times 
   are then integer multiples of $g t_{rev} = \pi \sqrt{20} \approx 14.05$.
%
% \noindent

   When the parameter $k^2$ does not satisfy \eq{param_k2} the time-dependence of
 $p_{\beta\alpha}(t)$ is 
   much more complicated. As long as $|\alpha| \neq 0$ and $|\beta| \neq 0$ 
   the probability in \eq{p_rewrite} will then never reach unity.
   Hence, there are no revivals.
   But as we can see from \fig{fig_p_zw} the probability $p_{\beta \alpha}(t)$
   nevertheless exhibits some peaks at e.g. $gt \approx 22$ and $gt \approx 41$ 
   (dotted curve). 
   These times correspond to  the condition $2 - \exp[A_0(t)] - \exp[A_0^*(t)] 
   \approx 0$, due to  the first factor in \eq{p_rewrite}. If this
condition is not satisfied  $p_{\beta\alpha}(t)$ will be exponentially suppressed
    when $\alpha$ and $\beta$ are large, i.e. when the photon number 
   is large.      
%\vspace{2cm}

%%%%%%%%%%%%%%%%%%%%%%%%%%%%%%%  qm_revivals.tex (end)  %%%%%%%%%%%%%%%%%%%%%%%%%%%%%%%%

%%%%%%%%%%%%%%%%%%%%%%%%%%%%%%%  non_CL_behaviour.tex (start)  %%%%%%%%%%%%%%%%%%%%%%%%%%%%%

\vspace{1cm}
%%%%%%%%%%%%%%%%%%%%%%%%%%%%%%%%%%%%%%%%%
\bc{
\section{Non-Classical Behavior}
}\ec
%%%%%%%%%%%%%%%%%%%%%%%%%%%%%%%%%%%%%%%%%
%  \noindent

  Our two-mode system exhibits quantum correlations. These correlations 
  can be described by various correlations functions. We find it convenient to
  consider the following well-defined second-order correlations function: 
\be \label{f}
 f(t) \equiv \sqrt{ \langle (a^{\da})^2 a^2 \rangle} \sqrt{ \langle (b^{\da})^2
             b^2 \rangle} -  \langle a^{\da} a ~ b^{\da} b \rangle~,
\ee                               
where e.g.

\be \label{moment}
   \langle a^{\da} a b^{\da} b \rangle \equiv \langle i |~ U^{\da}(t)~
         a^{\da} a~ b^{\da} b ~ U(t)~ | i \rangle ~.
\ee
\noindent
   Since the expectation-values in \eq{f} are normal-ordered
   we can always write these moments in terms of a Glauber-Sudarshan $P$ 
   representation, e.g. 
\be
\label{GlauberP}
\langle a^{\da} a b^{\da} b \rangle = \int d^2 \alpha 
   \int d^2 \beta ~|\alpha|^2 ~|\beta|^2~ P(\alpha,\beta;t)~~~.
\ee
   If $P(\alpha,\beta;t)$ is real and positive, then the moments in \eq{f} are 
   defining inner products. Hence, the Cauchy-Schwarz inequality implies 
   the $f(t)\geq0$. 
   States described by such a function admit a classical interpretation 
   where the complex field amplitudes $\alpha$ and $\beta$ may be treated as
   stochastic random variables with the joint probability distribution 
   $P(\alpha,\beta;t)$.
   If, however, $f(t) < 0$ then $P(\alpha,\beta;t)$ is not real and positive
   everywhere. 
   In such a situation the correlations cannot be described classically
   since $P(\alpha,\beta;t)$ cannot be interpreted as a 
   classical probability distribution. Hence, the sign of $f(t)$
   reflects the classical or quantum nature of the correlations
   between 
the modes.
%
%
%

%   \noindent

   Another convenient way to characterize a non-classical state for
   single modes is  the
   Mandel quality factor $Q_a$ defined by \cite{Mandel&Wolf&95}
\be \label{Q_a}
   Q_a(t) = \frac{\langle ( \Delta n_a(t) )^2 \rangle - 
            \langle n_a(t) \rangle}{ \langle n_a(t) \rangle  }~,
\ee
    where $n_a(t) \equiv a^{\da}(t)a(t)$ and $(\Delta {\cal O}(t))^2 \equiv 
    Var[{\cal O}(t)] \equiv
    \langle {\cal O}^2(t) \rangle - \langle {\cal O}(t) \rangle ^2$
    for an observable ${\cal O}$.
    The quality factor is constructed in such a way that $Q_a=0$ when the 
    statistics describing the isolated $a$-quanta is Poissonian. 
    If the $Q_a$ factor is negative the state is quantum mechanical,
    without any classical analog. In this case the statistics is sub-Poissonian. The
    smallest value $-1$ is obtained for a Fock state.   
    On the other hand, when $Q_a$ is positive the statistics is super-Poissonian
    and the isolated $a$-mode can be described in classical terms. 
    A corresponding formula as \eq{Q_a} holds for the $b$-quanta.
\vspace{2mm}
%%%%%%%%%%%%%%%%%%%%%%%%%%%%%%%%%%%%%%%%%
\bc{
\subsection{Fock States}
}\ec
%%%%%%%%%%%%%%%%%%%%%%%%%%%%%%%%%%%%%%%%%
%
%
%    \noindent

\begin{figure}[t]
\unitlength=1mm
\begin{picture}(160,80)(0,0)
\includegraphics{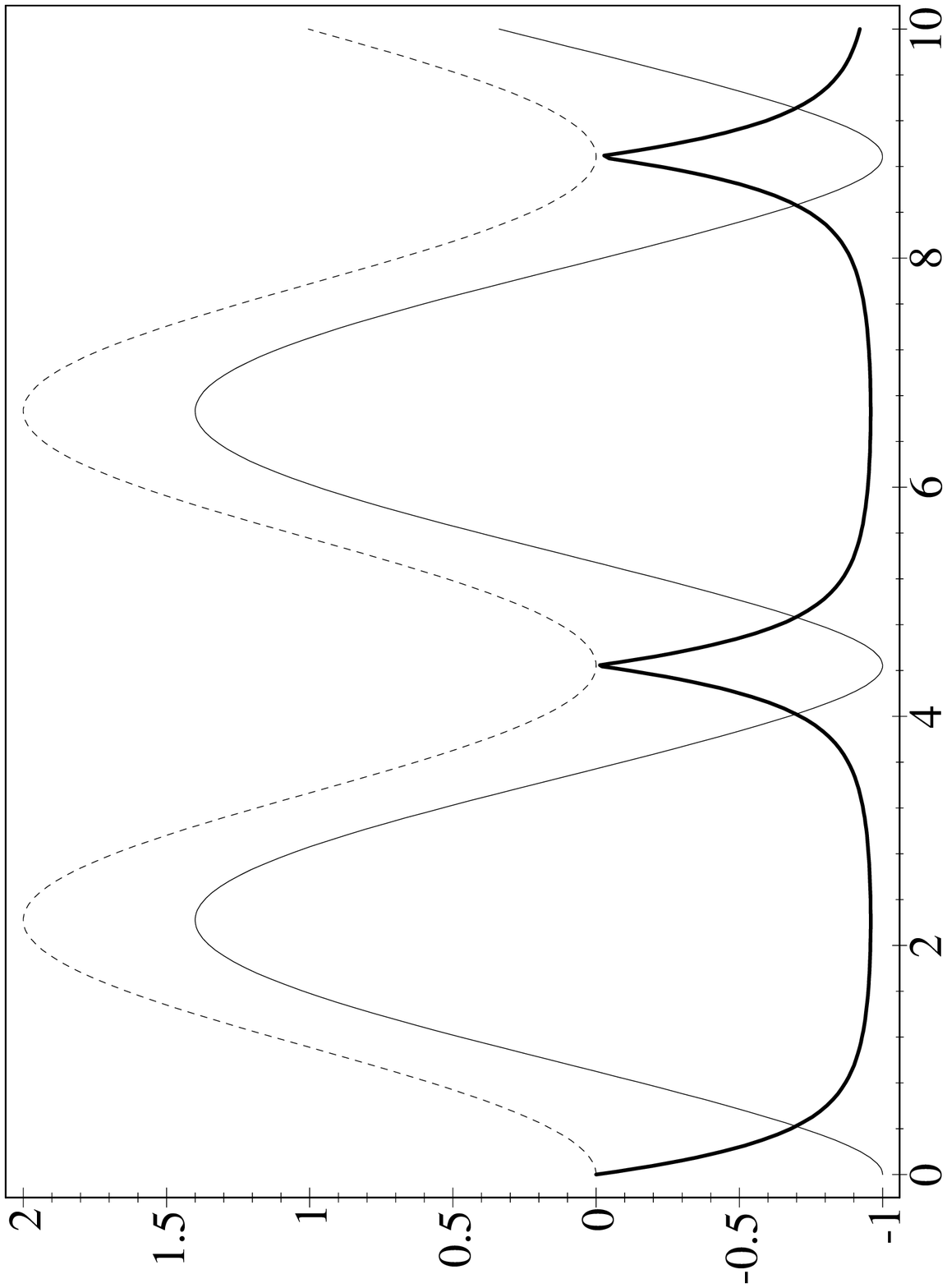}
   \put(78,0){\normalsize  \boldmath$gt$}
   \put(0,40){\normalsize  \boldmath$Q_a(t)$}
   \put(0,55){\normalsize  \boldmath$ F(t)$}
\end{picture}
\vspace{2mm}
\figcap{ The Mandel factor $Q_a(t)$ when the system is in the Fock state
  $|r \rangle _a \otimes |s \rangle _b$ and when 
  $k^2=1.5$. The thin solid curve correspond to
  $r=1$, $s=0$ and the dotted curve correspond to $r=0$, $s=1$.
  The thick solid curve shows the corresponding $F(t)$-function.
  \label{q_q_fock_1_5} }
\end{figure}

    Let us now consider the situation with initial  Fock 
    states $|r \rangle _a \otimes |s \rangle  _b \equiv |r,s \rangle$. 
    The Mandel factor in this case is given by
\be \label{Q_a_fock}
  Q_a(t) = \frac{n_0(t)~2rs + 
            n_0^2(t)~[2rs + r + s + 1] - r}{r + n_0(t)~[r + s + 1]}~,
\ee
    where $n_0(t)$ is as in \eq{z_label}. 
    If $r=0$ then $Q_a(0)=0$, independent of $s$, see e.g. \fig{q_q_fock_1_5}
    (dotted curve). On the other hand, if $r\neq0$ then $Q_a(0)=-1$, also
    independent of $s$, see e.g \fig{q_q_fock_1_5} (thin solid curve).
    The physics when $t>0$   depends strongly on the value of $k^2$. 
    If $k^2<1$ the Mandel factor increase exponentially with time,
    which leads to super-Poissonian statistics of the
     $a$-quanta.
    If, on the other hand, $k^2>1$ then the $Q_a(t)$-function will
    have an oscillatory behavior.
    The maximum value of $Q_a(t)$, when $r\neq0$ and $s=0$, is 
    $Q_a^{max} = [1+1/r-(k^2-1)^2]/[(k^2-1)^2 + (k^2-1)(1+1/r)]$. 
    If $r=0$ and $s\neq0$ then $Q_a^{max} = 1/(k^2-1)$.
    The minimum value of $Q_a(t)$ is $0$ or $-1$, depending on the value of 
    $r$ as discussed above (c.f. \fig{q_q_fock_1_5}).
%
%

%   \noindent

    The $f(t)$-function when the system is initially in the Fock state
    $|r,s \rangle$ is given by 
\bea \label{f_fock}
\nonumber f(t)~~ = 
      && x^2(t)~\sqrt{r(r-1) + 4r(s+1)y(t) +  (s+1)(s+2)y^2(t) } \\ \nonumber 
      && \times \sqrt{s(s-1) + 4s(r+1)y(t) + (r+1)(r+2)y^2(t) } \\
      && -~ x^2(t)~[~ rs + (r+1)(s+1)y^2(t) \\ \nonumber 
      && +~\left( rs + r(r+1) + s(s+1) + 
         (r+1)(s+1) \right )y(t) ~  ]~, \nonumber
\eea 
    where $x(t)$ is given by \eq{x_label} and $y(t)=1-1/x(t)$ as in
\eq{y_label}. Notice that $f(t)$ is 
    symmetric with respect to exchange of $r$ and $s$ as it should.
%
%
%    \noindent

    Let us now, for reasons of convenience, consider the 
    correlation function $F(t) \equiv f(t)/\sqrt{\langle n_a(t) \rangle 
    \langle n_b(t) \rangle }$
    instead of $f(t)$ itself. 
    \fig{q_q_fock_1_5} shows $F(t)$ as well as the corresponding
    quality factors for $k^2 = 1.5$ and $r=1, s=0$ or $r=0, s=1$. 
The $F(t)$-function in this figure is negative for 
    all times, except at the revival times where it is zero, and 
     the two-mode down-converted light therefore
exhibits quantum mechanical correlations in this case. 
    Since the Fock state is a typical quantum mechanical state 
    this result is as expected. Nevertheless, the quality factor
    oscillates between $0$ and $2$ (dotted curve) indicating that 
    the statistics of the isolated $a$-system is super-Poissonian.
\begin{figure}[t]
\unitlength=1mm
\begin{picture}(160,80)(0,0)
\includegraphics{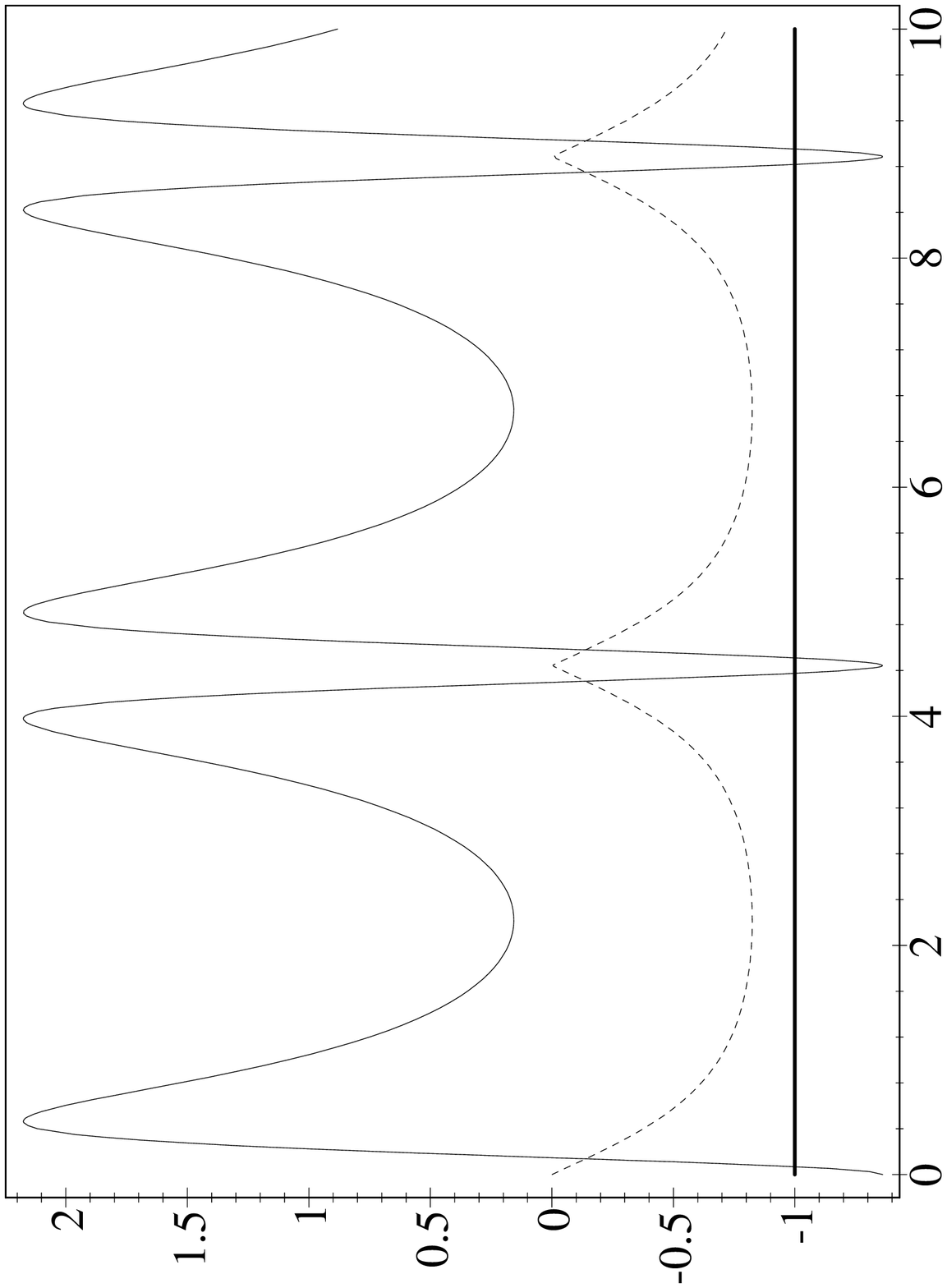}
   \put(78,-5){\normalsize  \boldmath$gt$}
   \put(0,50){\normalsize   \boldmath$F(t)$}
\end{picture}
\vspace{2mm}
\figcap{The normalized function $F(t) \equiv f(t)/\sqrt{n_a(t) n_b(t)}$ when 
        $k^2=1.5$.
        The thin solid cure shows $F(t)$ when the system is initially
        a Fock state $|r \rangle_a \otimes |s \rangle _b$
        with $r=50$, $s=10$ and the dotted curve when 
        $r=50$, $s=0$. The thick solid curve corresponds to arbitrary $r=s$.

\label{f_f_fock} }
\end{figure}
%
%
%    \noindent
        It is, however,  possible to make $F(t)$ positive in 
    spite of the fact that  
    the Fock state is a typical quantum mechanical state.
    This can be done by choosing $r \gg s \neq0$
    or vise versa (e.g. $r=50$ and $s=10$ as in \fig{f_f_fock} (thin solid 
    curve)). If $r=s$ then $F(t)=-1$ for all possible values of $k^2$, 
    c.f. the thick line in \fig{f_f_fock}.
    If $s=0$ then $F(t)\leq0$ also for all $k^2$. 
\begin{figure}[t]
\unitlength=1mm

\begin{picture}(160,80)(0,0)
\includegraphics{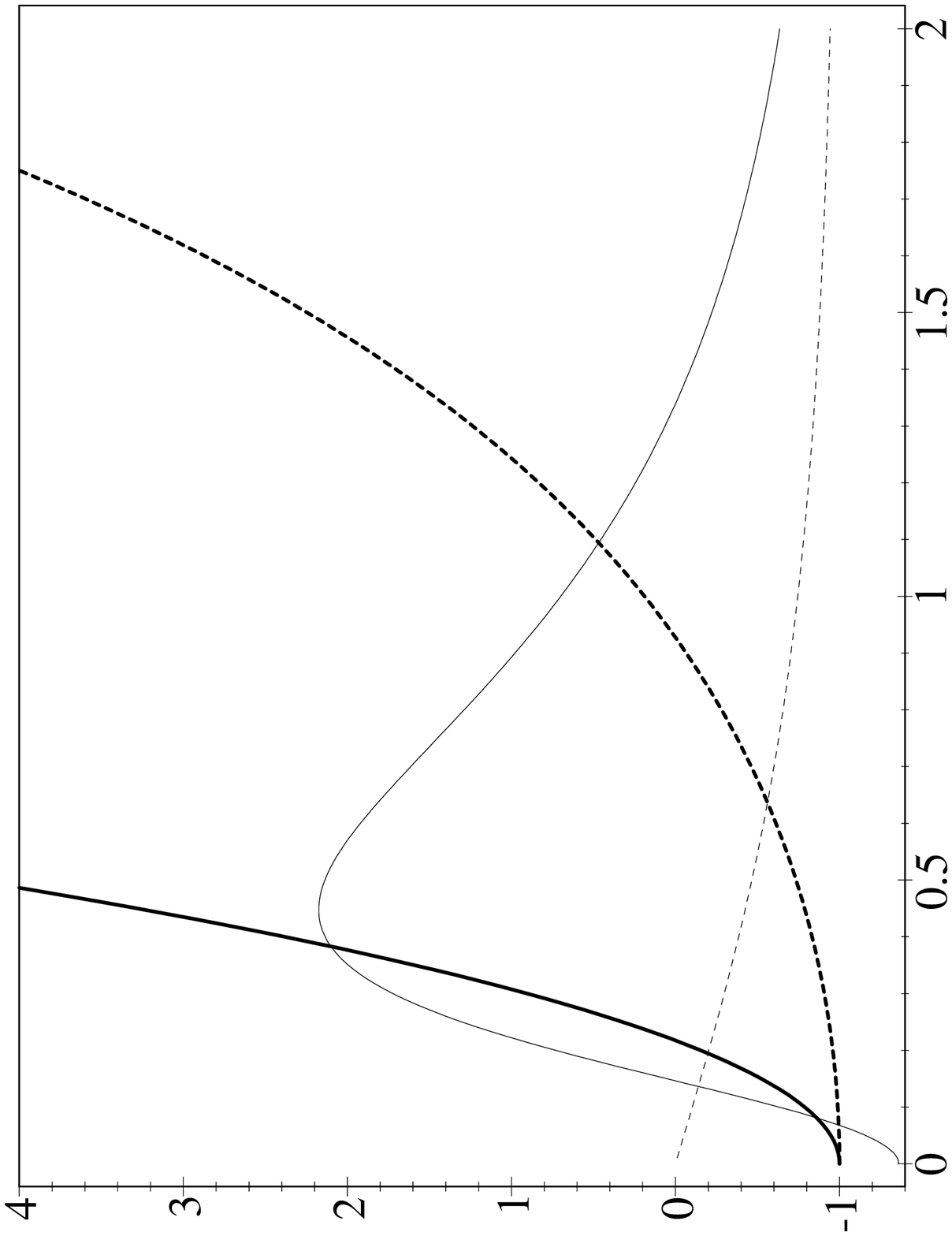}
   \put(78,-5){\normalsize  \boldmath$gt$}
   \put(0,55){\normalsize \boldmath$F(t)$}
   \put(0,35){\normalsize \boldmath$Q_a(t)$}
\end{picture}
\vspace{2mm}
\figcap{ All curves correspond to $k^2=0.5$. The thin solid curve shows 
         $F(t)$ when the two-mode system is initially 
         in the Fock state $|r \rangle _a \otimes |s \rangle  _b \equiv
         |r,s \rangle$ with  $r=50$, $s=10$. The thick solid curve is the
         corresponding $Q_a(t)$-factor. The thin dotted curve shows
         $F(t)$ in the
         case $r=50$, $s=0$ and the thick dotted curve is the corresponding
         $Q_a(t)$-factor. 

\label{f_peak_fock} }
\end{figure}
\noindent
    \fig{f_peak_fock} shows the typical behavior when $k^2<1$.  
    As we can see, $F(t)$ approaches $-1$ for sufficiently large times.
    Furthermore, the thin dotted curve in \fig{f_peak_fock} is negative for all times, 
    but the thin solid curve has a positive peak at small times ($0.1\lesssim gt 
    \lesssim 1.3$).
    In this short period of time the correlation between the $a$- and
    $b$-mode can be described in classical terms. 
    For all other times, when $F(t)$ is negative, the 
    down-converted quanta exhibits quantum mechanical correlations.
    Furthermore, the corresponding Mandel factor starts at $-1$ or $0$ 
    (depending on the value of $r$) and increase exponentially with
    time.

%    \noindent

    As mentioned earlier, in the case $k^2<1$, the isolated $a$- or $b$-mode
    describes a system with a temperature that approaches infinity in
    the large  $gt$ limit. It is therefore expected that the 
    statistics become super-Poissonian in this limit.
\vspace{5mm}
%%%%%%%%%%%%%%%%%%%%%%%%%%%%%%%%%%%%%%%%%
\bc{
\subsection{Coherent States}
}\ec
%%%%%%%%%%%%%%%%%%%%%%%%%%%%%%%%%%%%%%%%%
%
\begin{figure}[t]
\unitlength=1mm
\begin{picture}(160,80)(0,0)
\includegraphics{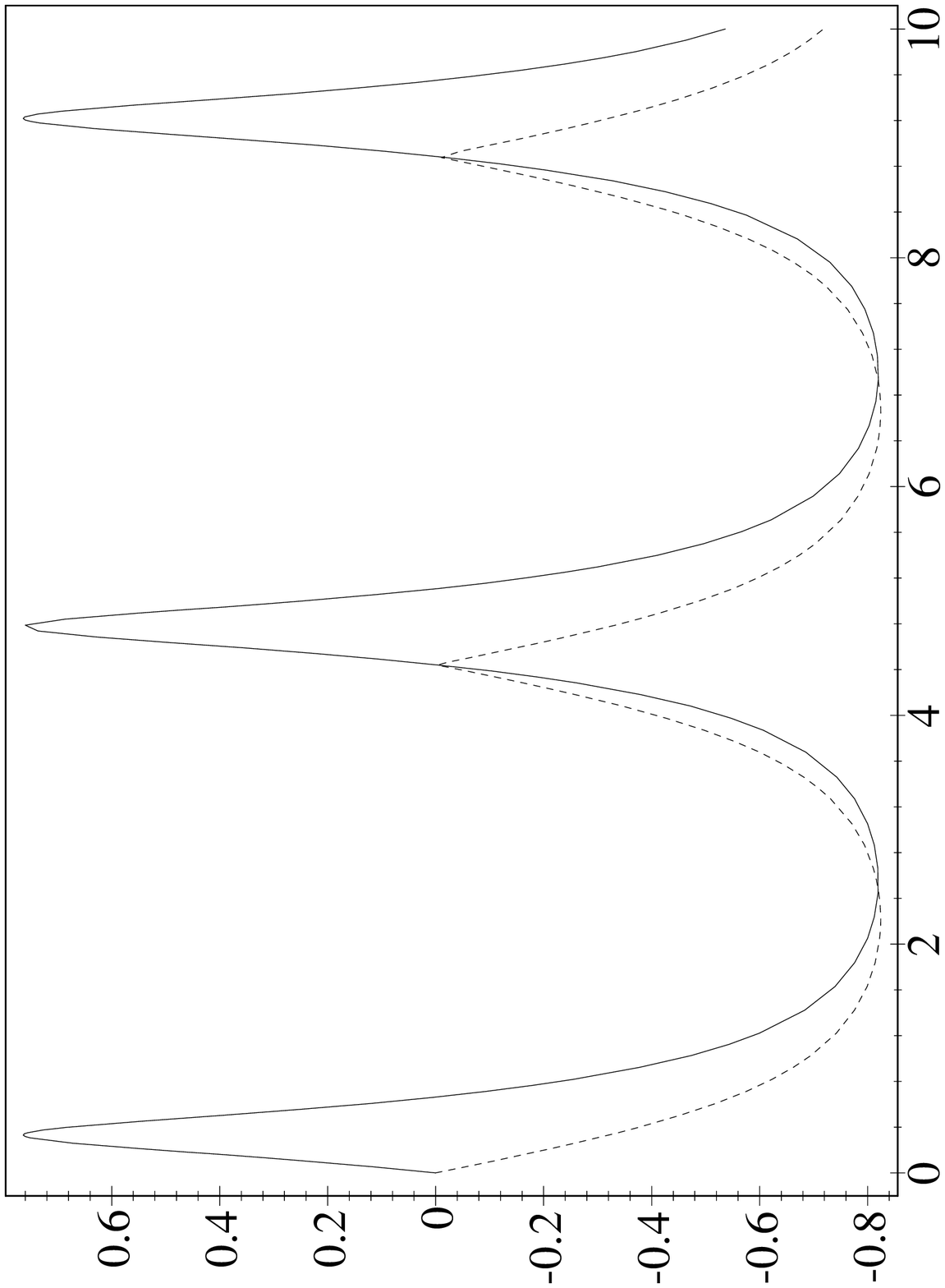}
   \put(78,-5){\normalsize  \boldmath$gt$}
   \put(0,50){\normalsize   \boldmath$F(t)$}
\end{picture}
\vspace{2mm}
\figcap{The correlation function $F(t) \equiv f(t)/\sqrt{\langle n_a(t) \rangle
        \langle n_b(t)} \rangle$ when $k^2=1.5$.
        The solid cure shows $F(t)$ when the system is initially
        a coherent state $|\alpha \rangle_a \otimes |\beta \rangle _b$
        with $\alpha=\sqrt{50}$, $\beta=\sqrt{10}$ and the dotted curve when 
        $\alpha=\sqrt{50}$, $\beta=0$.
\label{F_coh_1_5} }
\end{figure}
%
%
%
%    \noindent

    Since the  quantum mechanical state that corresponds
to a classical description is the coherent
    state, we  expect classical-like behavior when
    considering initial and final coherent states. 
\fig{F_coh_1_5} shows the function $F(t)$
    for the initial
    coherent state $|\alpha \rangle _a \otimes | \beta \rangle _b \equiv
    |\alpha, \beta \rangle$ when $k^2=1.5$. 
    For reasons of comparison we have chosen  the expectation-value of the
    photon number to be the same as in the Fock case in \fig{f_f_fock}.  
    We see that for  coherent states $F(t)$ is mostly negative 
    (solid curve in \fig{F_coh_1_5}) while for Fock states $F(t)$ is
    mostly positive (solid curve in \fig{f_f_fock}). 
    This illustrates the surprising fact that the correlation between the modes
    can be even more quantum mechanical for  coherent states  
    as compared to initial Fock states with the same average number of 
    photons.

%%%%%%%%%%%%%%%%%%%%%%%%%%%%%%%  non_CL_behaviour.tex (end)  %%%%%%%%%%%%%%%%%%%%%%%%%%%%%%

%%%%%%%%%%%%%%%%%%%%%%%%%%%%%%%  squeezing.tex (start)  %%%%%%%%%%%%%%%%%%%%%%%%%%%%%%
%
\vspace{1cm}
%
%%%%%%%%%%%%%%%%%%%%%%%%%%%%%%%%%%%%%%%%%
\bc{
\section{Squeezing and Detuning}
}\ec
%%%%%%%%%%%%%%%%%%%%%%%%%%%%%%%%%%%%%%%%%
%
\vspace{5mm}
%   \noindent

   In this section we will consider the influence of detuning on squeezing. 
   The squeezing will be defined with respect to the phase-dependent 
quadrature 
   amplitude $X_{\theta}(t)$ (see e.g. Ref.\cite{Walls&Milburn&95}):
\be \label{X_theta}
 X_{\theta}(t) = \frac{1}{\sqrt 2} \left \{ a(t)e^{i\theta t} + b(t)
               e^{-i\theta t} + h.c.  \right \}~.
\ee
   A number of schemes to make quadrature phase measurements have been 
   discussed in the literature (see e.g. Ref.\cite{Walls&Milburn&95}). 
   These schemes involve homodyning the signal field with a
   reference signal known as the local oscillator. In this case
   $\theta$ is the phase of the local oscillator.
%
%\noindent
%
   The variables $X_0(t)$ and $X_{\pi/2}(t)$ are canonical, i.e. they
obey the commutation relation
\be\label{com_rel_X}
[X_0,X_{\pi/2}]= -2 i ~.
\ee
%
%
%   \noindent
   The corresponding uncertainty product  therefore is
\be \label{unc_prcip}
 \Delta X_0 \Delta X_{\pi/2} \geq 1 ~~~.
\ee
We now consider squeezing in terms of the observable $X_{\theta}(t)$ for initial
Fock states or coherent states.
\vspace{5mm}
%
%%%%%%%%%%%%%%%%%%%%%%%%%%%%%%%%%%%%%%%%%
\bc{
\subsection{The Fock State}
}\ec
%%%%%%%%%%%%%%%%%%%%%%%%%%%%%%%%%%%%%%%%%
%
 %   \noindent

    If the system initially is in the Fock state $|r,s \rangle$
    the variance in $X_{\theta}(t)$ is expressed by
\be \label{var_X_fock}
 Var[X_{\theta}(t)] = |T_{\theta}(t)|^2 \left ( r+s+1 \right )~~~,
\ee
where 
\bea \label{T_2} 
 && |T_{\theta}(t)|^2 \equiv 
                    |e^{\displaystyle -A_0^*(t) + i\theta } - 
                    A_-(t)~e^{\displaystyle -A_0(t) - i\theta }|^2 \nonumber \\
&&  =~x(t) \left[ 1 + y(t)~  
                     - ~ 2 \left (~ \cos(\Omega t - 2\theta) G(t)
                                  + \sin(\Omega t - 2\theta) H(t)
                                     ~ \right) ~\right]~~~.
\eea
Here we have defined the functions
\be
G(t) =
\left\{ 
\begin{array} {rcl} 
   ~~\sqrt{k^2-1}~\frac{\displaystyle  \tan(g t \sqrt{k^2-1})~ [ 1 - \tanh^2\delta]} 
      {\displaystyle \tan^2(g t \sqrt{k^2-1}) + \tanh^2\delta}  &~,~~\mbox{if}&  k^2 \geq 1  \\ \\
   ~~\sqrt{1-k^2}~\frac{\displaystyle \tanh(g t \sqrt{1-k^2})~ [ 1 + \tan^2\gamma]} 
      {\displaystyle 1 + \tanh^2(g t \sqrt{1-k^2}) \tan^2\gamma}           
       &~,~~\mbox{if}&  k^2 \leq 1
\end{array}
\right.
\ee
and
\be
H(t) =
\left\{ 
\begin{array} {rcl} 
   ~~\sqrt{k^2-1}~ [   \coth\delta -
     \frac{\displaystyle  \tanh\delta~[ 1 + \tan^2(gt\sqrt{k^2-1)}]} 
     {\displaystyle \tan^2(g t \sqrt{k^2-1}) + \tanh^2\delta}] 
     &~,~~\mbox{if}&  k^2 \geq 1  \\ \\
   ~~\sqrt{1-k^2}~ [ \tan\gamma - 
     \frac{\displaystyle \tan\gamma~[ 1 - \tanh^2(gt\sqrt{1-k^2})]} 
     {\displaystyle 1 +\tanh^2(g t \sqrt{1-k^2})\tan^2\gamma}]
      &~,~~\mbox{if}&  k^2 \leq 1~~~.
\end{array}
\right.
\ee
    \noindent
    $x(t)$ and $y(t)$ are given by \eq{x_label} and \eq{y_label}, 
    respectively. The expressions for $\tan\gamma$ and 
    $\coth\delta$ are, furthermore,  given by \eq{tan_alpha} and  \eq{coth_beta}.
%
%    \noindent
    The special case when $k^2=0$ the variance in \eq{var_X_fock} reduces to
\be
 Var[X_0(t)] = e^{-2gt}(r+s+1)~~~,
\ee
and
\be
 Var[X_{\pi/2}(t)] = e^{2gt}(r+s+1)~~~,
\ee
     in agreement with well known results (see e.g. \cite{Walls&Milburn&95}).
    We immediately see that this is a minimum-uncertainty state when $r=s=0$.
    Therefore, changing the phase of the local oscillator, $\theta$, by $\pi/2$ 
    enables one to move from enhanced to squeezed quadrature phase fluctuations.
%
    
%\noindent

 By inspection of the actual expressions for $A_+(t)$ and $A_-(t)$ in 
    \eq{A_bigger_1} we see that \eq{T_2} is periodic when $k^2$ 
    satisfies \eq{param_k2}. The corresponding revival time is therefore
    again given by \eq{coh_rev_time}. 
\begin{figure}[t]
\unitlength=1mm
\begin{picture}(160,80)(0,0)

\includegraphics{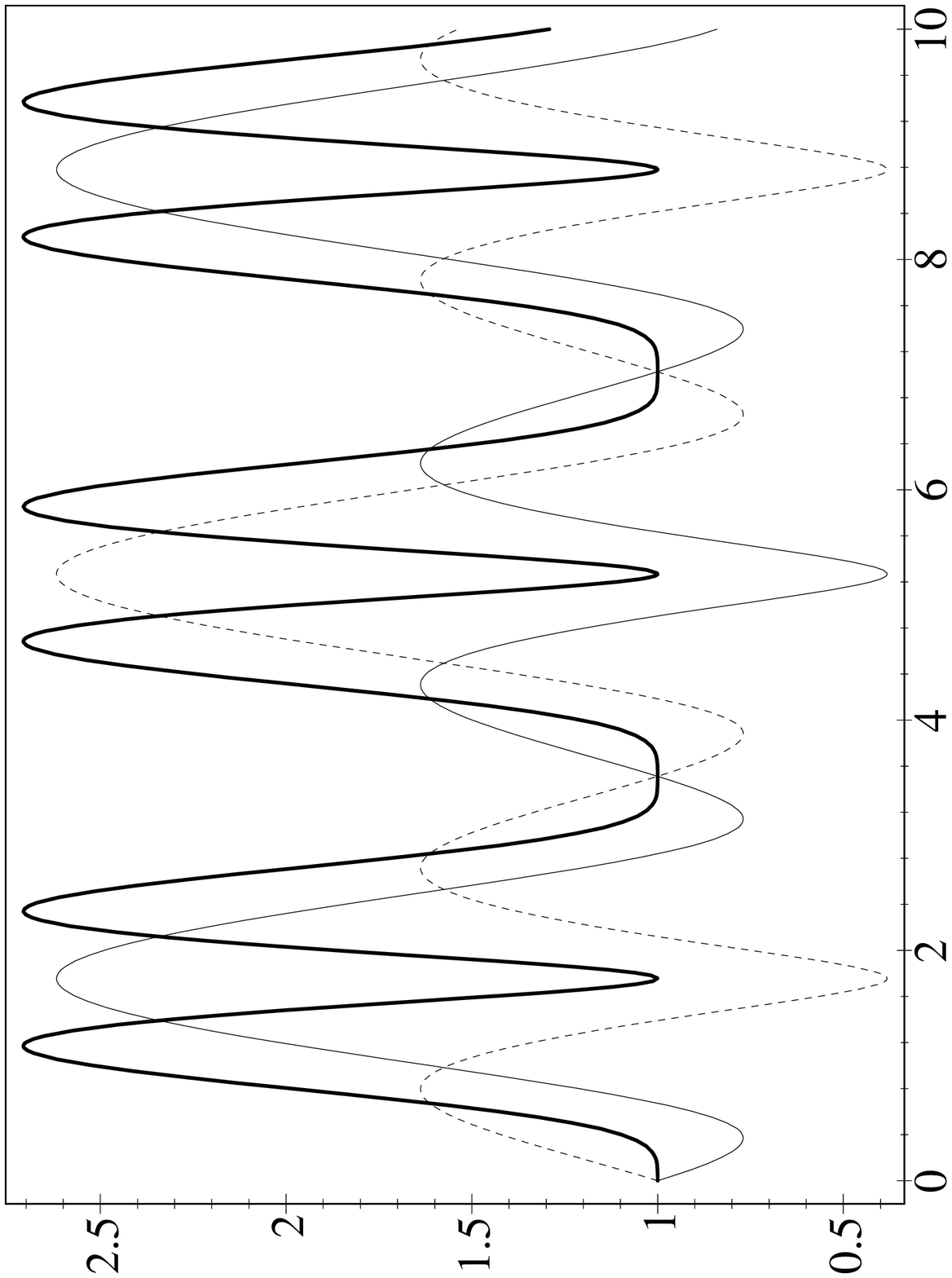}
   \put(76,-5){\normalsize  \boldmath$gt$}
   \put(-5,70){\normalsize  \boldmath$\Delta X_0$}
   \put(-5,55){\normalsize  \boldmath$\Delta X_{\pi/2}$}
   \put(-10,40){\normalsize  \boldmath$\Delta X_0   \Delta X_{\pi/2}$ }
\end{picture}
\vspace{2mm}
\figcap{ The thin solid curve shows $\Delta X_0(t)$, the dotted curve
         shows $\Delta X_{\pi/2}(t)$. The thick solid curve is
         the corresponding uncertainty product. All curves correspond
         to the vacuum case and $k^2=9/5$. 
\label{x_x_9_5} }
\end{figure}
%
%
%    \noindent
    \fig{x_x_9_5} shows the uncertainties
    $\Delta X_0(t) \equiv \sqrt{Var[X_0(t)]}$ (thin solid curve) and 
    $\Delta X_{\pi/2}(t) \equiv \sqrt{Var[X_{\pi/2}(t)]}$ (dotted curve)
    when the system is initially in vacuum for $k^2 = 9/5$. These curves oscillate 
    with the period $ gt_{rev} = \pi \sqrt{5} \approx 7.02$, where we can
    choose $p=2$ and $n=3$. However,
    the period of the uncertainty product is just half the period of the 
    individual uncertainties, i.e. uncertainty product oscillate with 
    period $ gt = \pi \sqrt{5}/2 \approx 3.51$. This is due to the fact
    that $Var[X_0(t+t_{rev}/2)] = Var[X_{\pi/2}(t)]$, i.e. the curve
    $Var[X_{\pi/2}(t)]$ is just shifted $t_{rev}/2$ relative to
    $Var[X_0(t)]$. 

    Due to revivals the uncertainty product is one at the revival times.
    The uncertainties $\Delta X_0(t)$ and $\Delta X_{\pi/2}(t)$ 
    are then equal and there is no squeezing. We now observe that we can write
    $\Delta X_0^2(t) = c(t) + d(t)$ and $\Delta X_{\pi/2}^2(t) = c(t) - d(t)$,
    i.e. the uncertainty product is $\Delta X_0^2(t)\Delta
    X_{\pi/2}^2(t)= c^2(t)-d^2(t)$, where the functions $c(t)$ and
    $d(t)$ can be obtained from \eq{T_2}. From this observation we
    infer that extremal values of $\Delta X_0(t)$ {\it and} $\Delta
    X_{\pi/2}(t)$
are associated with extremal values of the uncertainty product. This
    fact is also clearly illustrated in  \fig{x_x_9_5}.
\begin{figure}[t]
\unitlength=1mm
\begin{picture}(160,80)(0,0)
\includegraphics{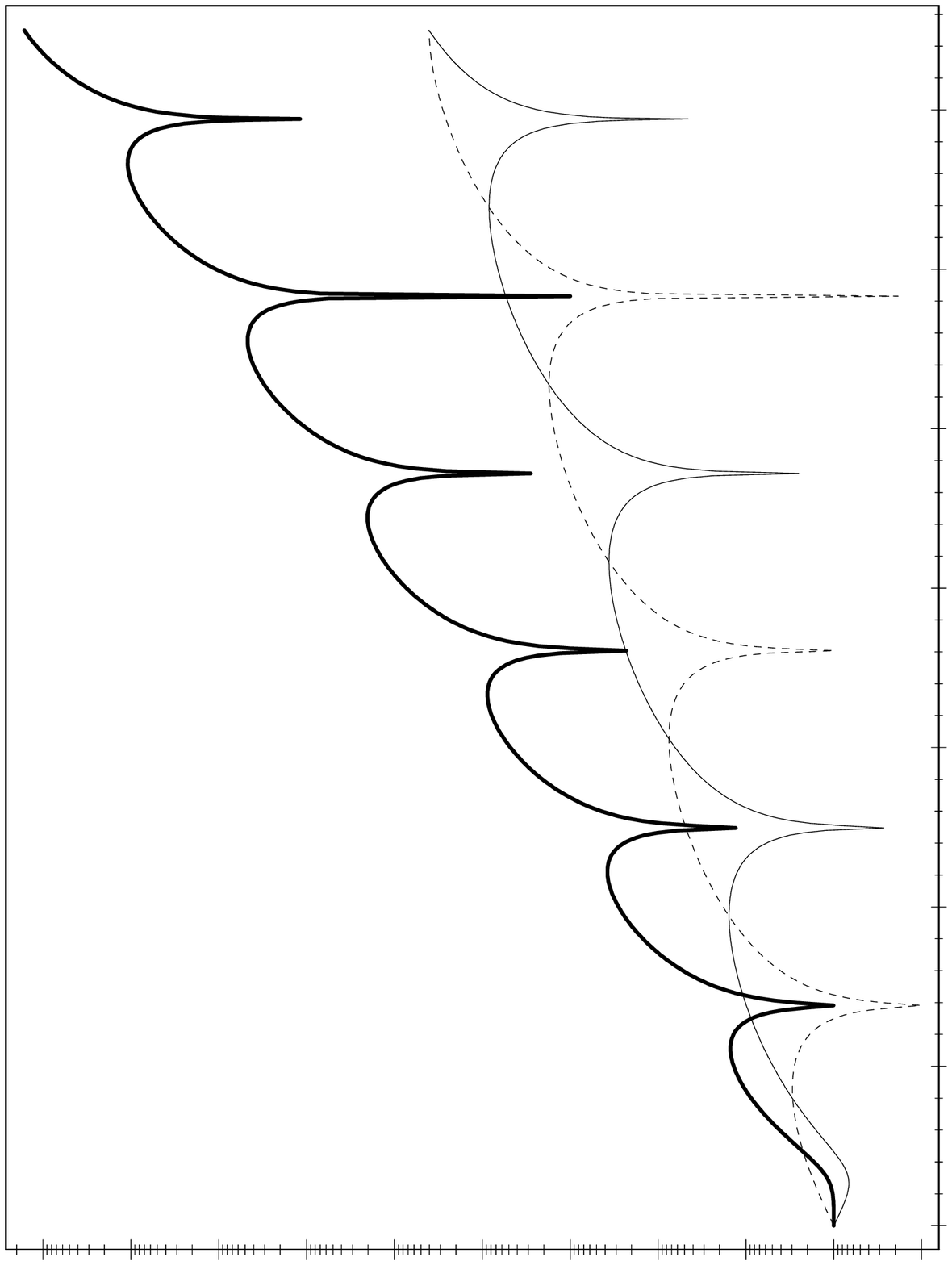}
       \put(20,-1){\normalsize $0$}
       \put(36,-1){\normalsize $2$}
       \put(52,-1){\normalsize $4$}
       \put(68,-1){\normalsize $6$}
        \put(82,-1){\normalsize $8$}
         \put(98,-1){\normalsize $10$}
          \put(114,-1){\normalsize $12$}
           \put(130,-1){\normalsize $14$}
       \put(10,92){\normalsize $10^9$  }
       \put(10,84){\normalsize $10^8$  }
       \put(10,74){\normalsize $10^7$  }
       \put(10,66){\normalsize $10^6$  }
       \put(10,57){\normalsize $10^5$  }
        \put(10,48){\normalsize $10^4$  }
         \put(10,40){\normalsize $10^3$  }
         \put(10,32){\normalsize $10^2$  }
         \put(10,24){\normalsize $10^1$  }
         \put(10,14){\normalsize $10^0$  }
          \put(10,6){\normalsize $10^{-1}$  }
   \put(75,-8){\normalsize  \boldmath$gt$}
   \put(30,70){\normalsize  \boldmath$\Delta X_0$} 
   \put(30,78){\normalsize  \boldmath$\Delta X_{\pi/2}$ }
   \put(30,86){\normalsize  \boldmath$\Delta X_0   \Delta X_{\pi/2}$}
\end{picture}
\vspace{2mm}
\figcap{ The thin solid curve shows $\Delta X_0(t)$, the dotted curve
         shows $\Delta X_{\pi/2}(t)$. The thick solid curve is
         the corresponding uncertainty product. All curves correspond
         to the vacuum case and $k^2=0.5$. 
\label{fig_log_9_5} }
\end{figure}
%
    
%\noindent

    Let us now study the case $k^2<1$. 
    \fig{fig_log_9_5} shows a logarithmic plot of the uncertainties $\Delta X_0(t)$ 
    and $\Delta X_{\pi/2}(t)$ and the corresponding  uncertainty product when the system 
    is initially in the vacuum state and $k^2=0.5$. 
    As we can see, 
    the uncertainties increase essentially at an exponentially rate as
    a function of time. However, the 
    curves also exhibit an oscillatory behavior with various local
    minima. 
These minima actually occur 
    when the time-derivative of \eq{T_2} is equal to zero. 
    When $gt\sqrt{1-k^2}\ll 1$ 
    the derivative of this equation is complicated and the various
    minima must be determined numerically.  
    When $gt\sqrt{1-k^2}\gg 1$ the
    periodicity for large $gt$ is determined by the the cosine- and 
    sine-functions in \eq{T_2}, 
    i.e. the local minima of $\Delta X_0(t)$ and $\Delta X_{\pi/2}(t)$ have a period 
    $gt = \pi/k$. The period of the uncertainty product is half of this period, 
    as discussed above.
%
%  almost squeezing 
%
\vspace{5mm}
%%%%%%%%%%%%%%%%%%%%%%%%%%%%%%%%%%%%%%%%%
\bc{
\subsection{The Coherent State}
}\ec
%%%%%%%%%%%%%%%%%%%%%%%%%%%%%%%%%%%%%%%%%
%
%
%    \noindent

    The coherent case, when the $a$-and $b$-mode are initially 
    in the coherent state $|\alpha,\beta \rangle$, the variance of $X_{\theta}(t)$ 
    is given by
\be \label{var_X_coh}
 Var[X_{\theta}(t)] = |T_{\theta}(t)|^2~,
\ee
     where again $|T_{\theta}(t)|^2$ is given by \eq{T_2}.
     The most striking feature of \eq{var_X_coh} is the fact that it 
     is independent of the initial values of $\alpha$ and $\beta$. 
     Moreover, the only
     difference between the Fock variance in \eq{var_X_fock} and the 
     coherent variance in \eq{var_X_coh} is the factor $r+s+1$.  
     The coherent state variances are therefore equal to the vacuum state variances.
 %

%%%%%%%%%%%%%%%%%%%%%%%%%%%%%%%  squeezing.tex (end)  %%%%%%%%%%%%%%%%%%%%%%%%%%%%%%

%%%%%%%%%%%%%%%%%%%%%%%%%%%%%%%%%%  stn.tex (strat)  %%%%%%%%%%%%%%%%%%%%%%%%%%%%%%%%%%%

                                         %
\vspace{1cm}      
%
%%%%%%%%%%%%%%%%%%%%%%%%%%%%%%%%%%%%%%%%%
\bc{
\section{Signal-to-Noise Ratio}
}\ec
%%%%%%%%%%%%%%%%%%%%%%%%%%%%%%%%%%%%%%%%%
 %   \noindent

    We define the signal-to-noise ratio $\rho_a (t)$ for the $a$-mode 
     by

\be \label{rho_ratio}
 \rho_a (t) = \frac{\langle ~ n_a(t) ~ \rangle}{\sqrt{\langle ~ Var[n_a(t)]~
 \rangle}} ~~,
\ee
    where $n_a(t) \equiv a^{\da}(t) a(t)$, see \eq{n_a_operator}. A
    similar definition for the $b$-mode holds.
    To get a good  signal we want such ratios to be as large as possible. 
    Let us suppose the system is initially in the Fock state $|r,s \rangle$. 
    The ratio $\rho_a(t)$ is then given by
\be \label{rho}
  \rho_a(t) = \frac{r + n_0(t)(r+s+1)}{\sqrt{n_0(t) + n_0^2(t) }}~
              \frac{1}{\sqrt{2rs+r+s+1}} ~~,
\ee
    where $n_0(t)$ is given by \eq{z_label}.
    The small and large time limits of $\rho_a(t)$ are

\vspace{3mm}
\begin{center}
\bea \label{grense}
 \rho_a(t) = \left\{ \begin{array}{ll}
                      {\displaystyle
                      \frac{r+s+1}{\sqrt{2rs+r+s+1}} }~
                      &~~,~~ \mbox{if $t \rightarrow \infty~,~k^2<1$~~,} \\ \\
                      {\displaystyle
                      \frac{r + (gt)^2 (r+s+1)}{gt \sqrt{2rs+r+s+1}} }~
                      &~~,~~ \mbox{if $t\rightarrow 0$~~,~~for all $k^2$~.}
                      \end{array}
               \right. 
\eea
\end{center}
%
%

%    \noindent

\begin{figure}[t]
\unitlength=1mm
\begin{picture}(160,80)(0,0)
\includegraphics{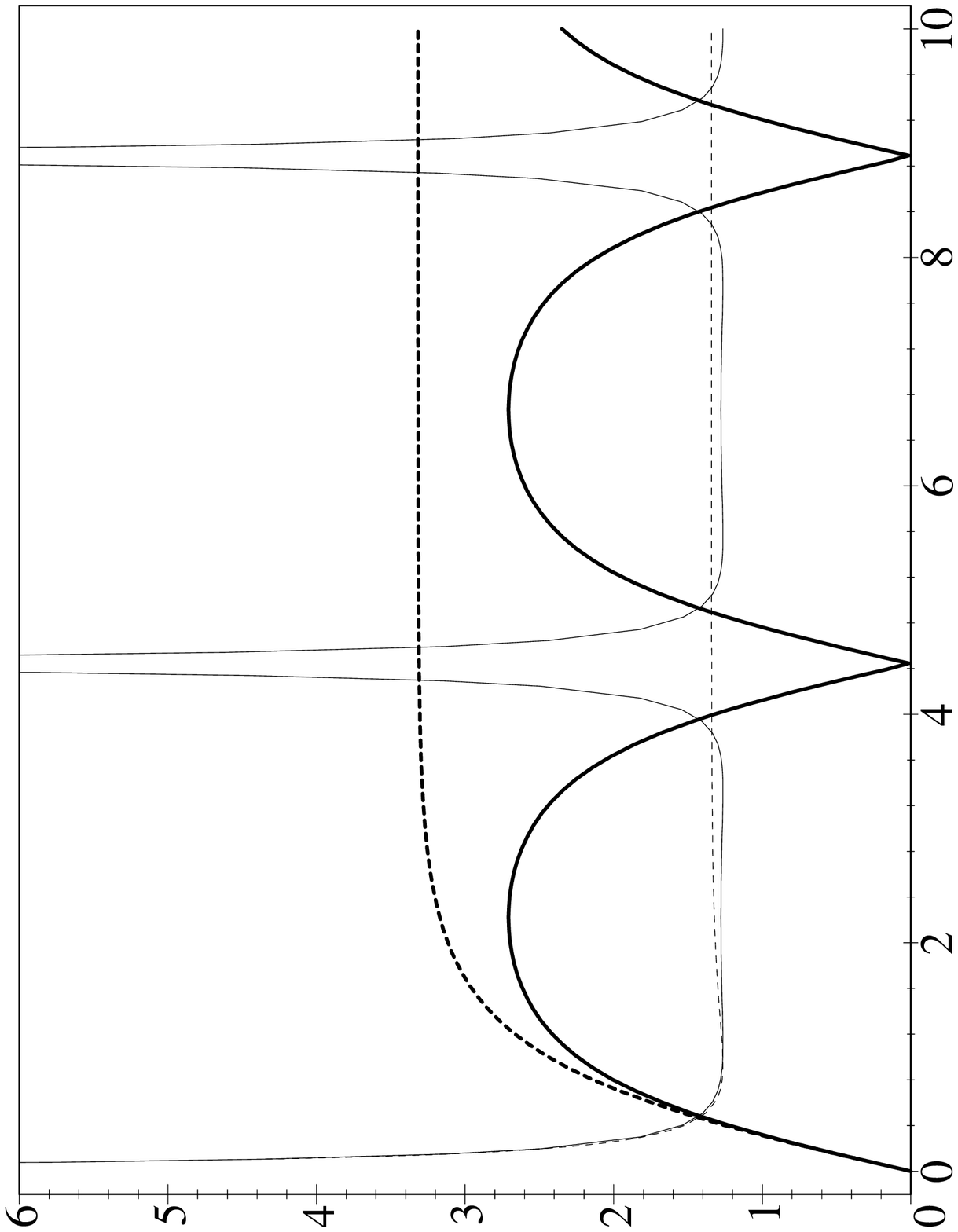}
   \put(78,-5){\normalsize  \boldmath$gt$}
   \put(0,50){\normalsize \boldmath$\rho_a(t)$ }
\end{picture}
\vspace{2mm}
\figcap{ The thin curves show $\rho_a(t)$ when $r=1$, $s=1$. The 
         thin solid curve corresponds to $k^2=1.5$ and the thin 
         dotted one corresponds to $k^2=0.5$.
         The thick curves show $\rho_a(t)$ when $r=0$, $s=10$. The 
         thick solid curve corresponds to $k^2=1.5$ and the thick 
         dotted one corresponds to $k^2=0.5$. The revival time is 
         $gt_{rev}=\sqrt{2}\pi \approx 4.44$. 
\label{rho_fig} }  
\end{figure}

    If $r=0$ we then have a well-defined limit for small times, i.e.
    $\rho_a(0)=0$ (see e.g. \fig{rho_fig}
    (thick curves)). On the other hand, if $r\neq0$ then $\rho_a(0)=\infty$  
    (see e.g. \fig{rho_fig} (thin curves)).
    This is due to the fact that a Fock state has no variance.
    The behavior of $\rho_a(t)$ at $t>0$ depends strongly on $k^2$.
    If $k^2<1$ the signal-to-noise ratio approaches the value as given in 
    \eq{grense}. If, on the other hand, $k^2>1$ the ratio $\rho_a(t)$ will 
    oscillate. The period of $\rho_a(t)$ is determined by \eq{z_label}, i.e. the 
    revival time is 

\be
gt_{rev}\sqrt{k^2-1} = n\pi~,
\ee 
    where $n$ is an positive integer.     
    The value of $\rho_a(t)$ at the revival time is $0$ or $\infty$, 
    depending on the value of $r$ as discussed above (see e.g.  
    \fig{rho_fig}). Therefore, we can achieve an infinite signal-to-noise 
    ratio at the revival times by choosing $r\neq0$.  
    
%\vspace{2cm}

%(When $r=0$, the maximum value of $\rho_a(t)$ is $\sqrt{s+1}/k$).

%    \noindent

    Another interesting feature of the signal-to-noise ratio 
  $\rho_a(t)$ is that it can be  approximatively constant between
  revival times, which also is illustrated  in  \fig{rho_fig}.
    This is to be compared with the oscillatory behavior of e.g.
    $p_{11}(t)$ ($p_{11}(t)$ in \fig{fig_p_nn_11_1_5} 
    is oscillating fast  for times where $\rho_a(t)$ is approximately constant).

%\vspace{2cm}
\begin{figure}[t]
\unitlength=1mm
\begin{picture}(160,80)(0,0)
\includegraphics{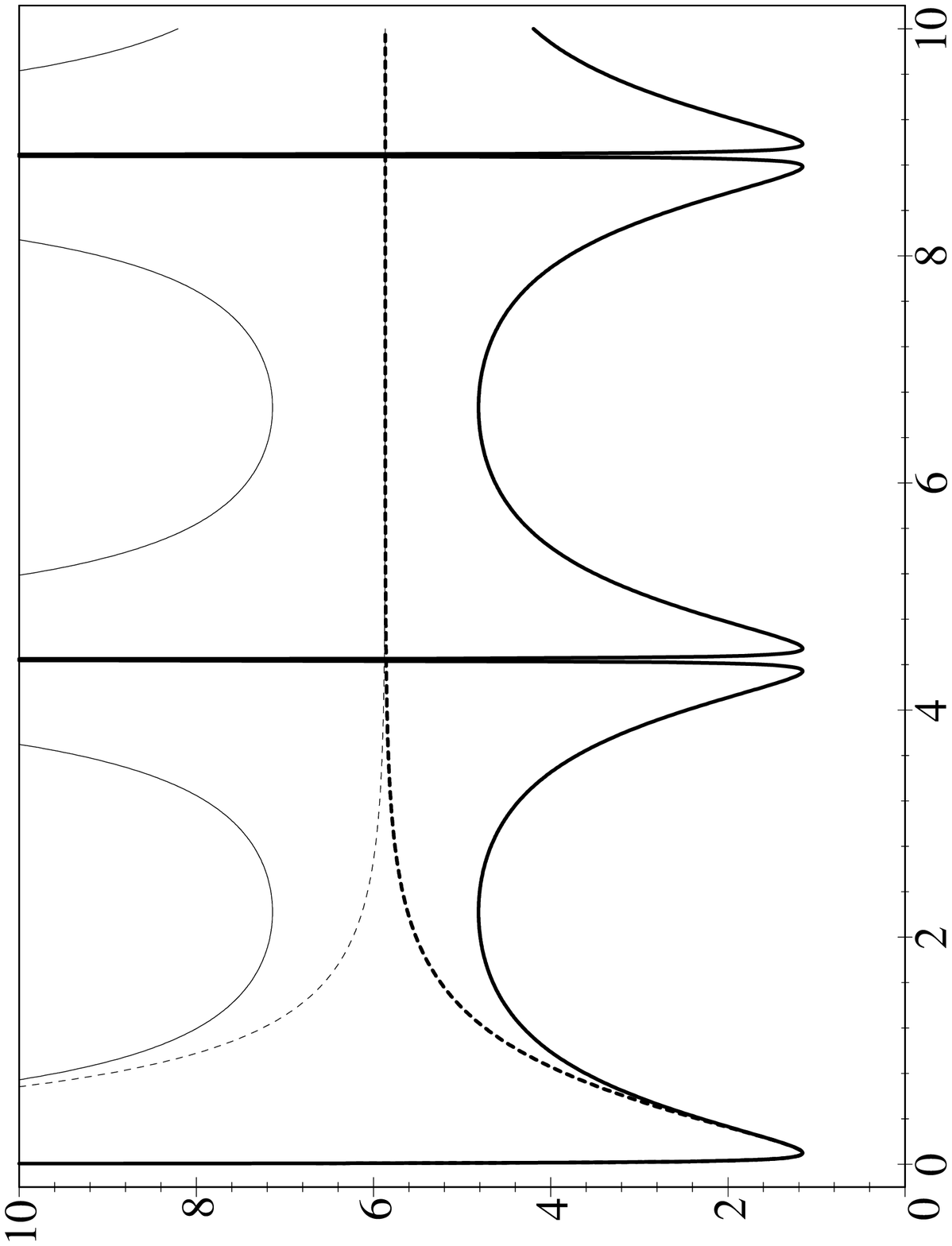}
   \put(78,0){\normalsize  \boldmath$gt$}
   \put(2,50){\normalsize \boldmath$\rho_a(t)$ }
\end{picture}
\vspace{2mm}
\figcap{ The thin curves show $\rho_a(t)$ when $r=100$, $s=1$. The 
         thin solid curve corresponds to $k^2=1.5$ with the
         global minimal value $\rho_a^{extr} = 152/\sqrt{452} 
         \approx 7.15$. The thin dotted one corresponds to $k^2=0.5$.
         The thick curves correspond to the case $r=1$, $s=100$.
         The thick solid curve show $\rho_a(t)$ when $k^2=1.5$
         with the local maximal value $\rho_a^{extr} = 205/(2\sqrt{452}) 
         \approx 4.82$. The thick dotted one corresponds to $k^2=0.5$.
         Both dotted curves  approach the value 
         $\rho_a = 102/\sqrt{302} \approx 5.87$.
\label{rho_extr_fig} }  
\end{figure}
    
%    \noindent

     Let us now study the
     extrema of the signal-to-noise ratio, i.e. when the time derivative of 
     \eq{rho} is zero. $\rho_a(t)$ has always an extremum when 
\be \label{t_extr}
  gt\sqrt{k^2-1} = n\pi/2~~~,
\ee 
    where $n$ is an odd integer. The value $\rho_a(t)$ at these particular times
    is
\be  \label{rho_extr} 
  \rho_a^{extr} = \frac{r k^2 +s+1}{\sqrt{k^2(2rs+r+s+1)}}~~~.
\ee
    If, in addition, $r$, $s$ and $k^2 (>1)$ satisfy the inequality 
\be \label{ulikhet}
  0 < \frac{r}{s-r+1} <  \frac{1}{k^2-1}~~~,
\ee
    the extremum in \eq{rho_extr} is a local maximum.  
    This inequality is e.g. satisfied when $r=1$, $s=100$ and $k^2=1.5$ 
    as shown in \fig{rho_extr_fig} (thick solid curve).
    The local maximum in this case is  $\rho_a^{extr} = 205/(2\sqrt{452}) 
    \approx 4.82$.
    Moreover, when \eq{ulikhet} is satisfied the ratio 
    $\rho_a(t)$ has two other extrema as well. 
    These extrema are global minima and occur when 

\be \label{t_min}
  gt\sqrt{k^2-1} = \pm \arcsin\left[\sqrt{(k^2-1)\frac{r}{s-r+1}}~\right] 
  + n_\pm \pi~~~,
\ee
    where $n_+=0,1,2,3,...$ and $n_-=1,2,3,...$. The particular value of 
    this minimum is 
\be \label{rho_min}
  \rho_a^{min} = 2 \sqrt{ \frac{1+1/s}{2+1/r+1/s+1/(rs)}}~~~,
\ee
    independent of the detuning parameter  $k^2$. The minimum value 
    corresponding to the thick solid curve in \fig{rho_extr_fig}
    is $\rho_a^{min} = 2 \sqrt{101/302} \approx 1.16$. 
   
%    \noindent

    If, on the other hand, \eq{ulikhet} is violated
    then $\rho_a^{extr}$ in \eq{rho_extr} is the only extremum.  It is a 
    global maximum when $r=0$ (see e.g. thick solid curve in \fig{rho_fig})
    and a global minimum when $r\neq0$ (see e.g. thin solid curve in 
    \fig{rho_extr_fig}). In both cases the global extremum is
    given by \eq{rho_extr}. 

%    \noindent 

    When the system is initially in a coherent state the signal-to-noise
    ratio is essentially the same as in the Fock case. The  essential
    difference 
    is that $\rho_a(t)$ is finite at the revival times since the
    variance is not zero.

 %   \noindent

\begin{figure}[t]
\unitlength=1mm
\begin{picture}(160,80)(0,0)
\includegraphics{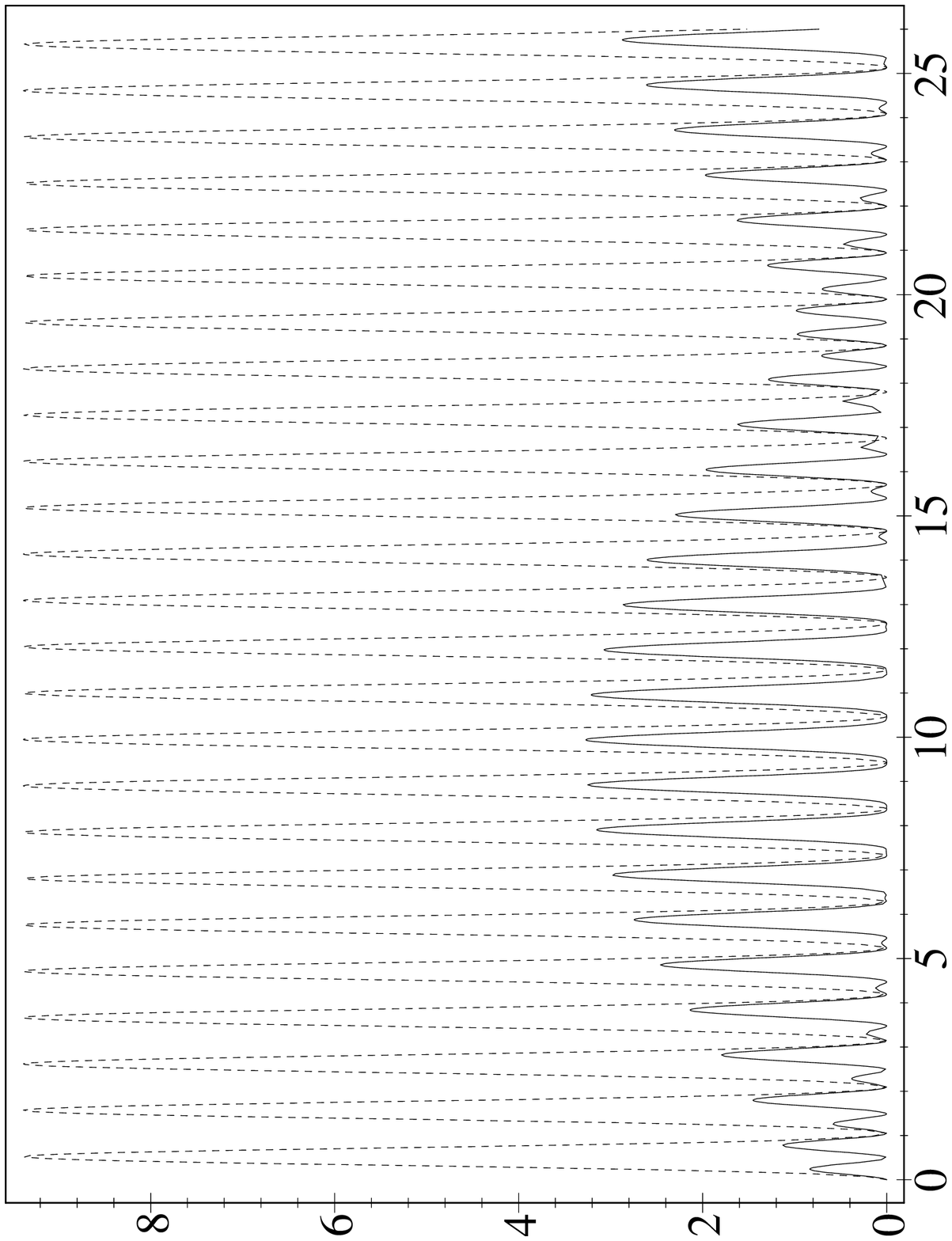}
   \put(76,-5){\normalsize \boldmath$gt$}
   \put(0,53){\normalsize \boldmath$\eta_a(t)$ }
\end{picture}
\vspace{2mm}
\figcap{The signal-to-noise ratio $\eta_a(t)$ when $k^2=10$.
        The system is initially in the coherent state $|\alpha,\beta \rangle$ 
        where $\alpha=0$ and $\beta=3$ (solid curve).
        The dotted curve is the corresponding Yuen-limit.
\label{eta_10_fig} }
\end{figure}

   An alternative definition of the signal-to-noise
    ratio can be obtained by considering the quadrature 
    operator 
\be \label{quadrature}
  X_a(t) = \frac{1}{\sqrt{2}}~ U_I^{\da}\left [ a + a^{\da} \right ]
           U_I ~~~,
\ee
   where $U_I$ is the time-evolution operator in \eq{U_int}.
   %The reason why we multiply these free time-evolution terms to the quadrature 
   %operator is to cancel the fast varying phase-term $exp(-i\omega_a t)$ in 
   %\eq{a_H_eq}.
   The signal-to-noise ratio is then defined by
\be \label{eta_def}
  \eta_a(t) = \frac{ \langle X_a(t) \rangle^2}{Var\left[X_a(t)\right]}~~~.
\ee
    From this definition we immediately see that $\eta_a(t)=0$ for a Fock state.
    \eq{eta_def} is, however,  non-trivial for  a coherent
    state $ |\alpha \rangle_a \otimes |\beta \rangle_b \equiv 
    |\alpha,\beta \rangle$. In this case the ratio $\eta_a(t)$ is
\be \label{eta}
  \eta_a(t) = \frac{ K(t) + 
                     x(t) \left\{~ |\alpha|^2 - 2 {\mbox Re}[\alpha \beta A_-(t)] + 
                     |\beta|^2 y(t)~\right\} }
                     {n_0(t)+ 1/2~}~~~,
\ee
where
\be \label{K_Yuen}
 K(t) \equiv  {\mbox Re}\left[e^{-2A_0^*(t)}(\alpha - \beta^*A_-^*(t))^2\right]~~~. 
\ee
    The functions $x(t)$, $y(t)$ and  $n_0(t)$ are given by \eq{x_label},
    \eq {y_label} and \eq{z_label}, respectively.
   The signal-to-noise ratio in \eq{eta} cannot take on any value:
   it has an upper bound. According to Yuen \cite{Yuen&76} the maximum 
   signal-to-noise ratio obtainable for the (two-photon) coherent 
   state $|\alpha,\beta \rangle$ cannot exceed the value 
   $4 \langle n_a(t)\rangle \left[ \langle n_a(t) \rangle+1 \right]$, i.e.
\be \label{eta_max}
\eta_a(t) \leq 4\langle n_a(t)\rangle \left[ \langle n_a(t) \rangle+1 \right]~,
\ee
   where $\langle n_a(t)\rangle = \langle \beta,\alpha 
   |a^{\da}(t)a(t)| \alpha,\beta \rangle$. It is natural to ask for
   what values of $\alpha$, $\beta$ and $k^2$ can we reach  this optimal Yuen-limit ?
   As an illustrative example 
   we choose a large $k^2$, e.g. $k^2=10$, and $\alpha=0$ and 
   $\beta=3$. The result of a numerical calculation is  shown in \fig{eta_10_fig}.
   As we can see from such a calculation, there
   is approximately a factor of $2.5$ in difference between the maximum 
   $\eta_a(t)$-value (solid curve) and the optimal Yuen-limit (dotted
   curve). 
   In other words, the signal is not very far from its optimal value.
   The price we have to pay to come this close to the 
   optimal Yuen-limit is the low value of $\eta_a(t)$ itself. 
   If we, on
    the other hand, choose a small $k^2$, e.g. $k^2=1.5$ (with the same 
   $\alpha$ and $\beta$), the physics is different.
   The maximum ratio 
   $\eta_a(t)$ is then ${\cal O}(10^1)$, which means that 
   the signal is strong as compared to the noise. In this case the
   Yuen-limit is, however, 
   ${\cal O}(10^3)$, so $\eta_a(t)$ is very far from the optimal
   Yuen-value.
   Numerical studies suggest that these observations are quite generic.

%\vspace{2cm}

%%%%%%%%%%%%%%%%%%%%%%%%%%%%%%%%%%  stn.tex (end)  %%%%%%%%%%%%%%%%%%%%%%%%%%%%%%%%%%%

%%%%%%%%%%%%%%%%%%%%%%%%%%%%%%%%%%  finalremarks.tex (start)  %%%%%%%%%%%%%%%%%%%%%%%%

%%%%%%%%%%%%%%%%%%%%%%%%%%%%%%%%%%%%%%%%%%%%%%%%%%%%%%%%%%%%%%%%%%%%%%%%%%%%%%%%%%%%%
%
\bc
{\section{Final Remarks \label{sec:finalremarks}}}
\ec
%
%%%%%%%%%%%%%%%%%%%%%%%%%%%%%%%%%%%%%%%%%%%%%%%%%%%%%%%%%%%%%%%%%%%%%%%%%%%%%%%%%%%%
%\noindent
  
By means of an, in general time-dependent,   Bogoliubov transformation
or what is called a two-mode
squeeze operator $S(z)$ in quantum optics \cite{Shum&86}, given by 
\be  
S(z) = \exp\left(z^*ab -za^{\da}b^{\da} \right)~~~,  
\ee 
where $z=r\exp(i\phi)$, one can diagonalize the Hamiltonian
\eq{schro_eq} instantaneously in terms of the canonical operators
\bea  
A(z)=S(z)aS^{\da}(z) &=& a\cosh r + b^{\da}\exp(i\phi)\sinh r ~~~,\nn \\  
B(z)=S(z)bS^{\da}(z) &=& b\cosh r + a^{\da}\exp(i\phi)\sinh r ~~~.  
\eea  
With the choice \eq{g_here} we e.g. find that
\be  
\label{eq:HOmega}  
H=\Omega_A A^\da (z) A(z) + \Omega_B B^\da (z)B(z) + \Omega_0  
\ee  
where 
\bea  
\Omega_A &=& \omega_{-} + \omega_{+}\sqrt{1-g^2/\omega_{+}^2}~~~, \nonumber \\  
\Omega_B &=& -\omega_{-} + \omega_{+}\sqrt{1-g^2/\omega_{+}^2}~~~,\nonumber \\  
\Omega_0 &=& \omega_{+}\sqrt{1-g^2/\omega_{+}^2}~~~,  
\eea 
and where $\omega_{\pm} = (\omega_{a}\pm \omega_{b})/2$. The squeeze
parameters are given by 
\be  
\cosh(2r) = 1/\sqrt{1-g^2/\omega_{+}^2}~~~,~~~  
\phi = \frac{\pi}{2} -\omega t ~~~.  
\ee  
The diagonalization procedure above is well-defined as long as $g^2 <
\omega_{+}^2$. If, on the other hand, $g^2 > \omega_{+}^2$ then the
instantaneous eigenvalues $\Omega_A$ and $\Omega_B$ become
complex. This indicates an instability in the system. As we have seen
in the present paper, one can explicitly
solve the time-dependent
Schr\"{o}dinger equation  in such a situation 
and, as we have seen above,  the physics  of the instability
actually manifest itself in an exponentially increasing particle, i.e. photon,  production.  
In the case $k^2 <1$ we have also
seen that the asymptotic time-limit of the reduced systems can be
described by thermal distributions. In general the distributions are
not identical (see e.g. Eqs.(\ref{reduced_a}) and (\ref{reduced_b})). 
But, nevertheless,
if the initial state is a pure state,
 the reduced entropies are the same
(see Refs.\cite{Araki&Lieb&70,Phoenix&Knight&88,Barnett&Phoenix&89})
and the entropy of the whole system is, of course, zero.

Some features of the simple model of parametric down-conversion as studied
in the present paper also enters in models considered in quantum
cosmology as e.g.  the
phenomena of non-adiabatic transitions (see
e.g. Ref.\cite{Massar&Parentani&98}). To the extent that such an analogy
holds we can therefore simulate some aspects of such models in the
laboratory by means of non-linear quantum optics.

%%%%%%%%%%%%%%%%%%%%%%%%%%%%%%%%%%  finalremarks.tex (end)  %%%%%%%%%%%%%%%%%%%%%%%%%%%%%%

%%%%%%%%%%%%%%%%%%%%%%%%%%%%%%%%%  acknowledgement.tex (start)  %%%%%%%%%%%%%%%%%%%%%%%%%%
%
%
\vspace{3mm}
\begin{center}
{\bf ACKNOWLEDGEMENT}
\end{center}
%
%\vspace{3mm}
%
We are greatful to K. Olaussen and A. Zeilinger for useful correspondence.
The research has been supported in part  by the Research Council of
Norway 
under 
contract no. 118948/410.
We are greateful to our colleagues at NTNU and in particular 
K. Olaussen for their interets in our work.
 \vspace{3mm}
%
%
%

%%%%%%%%%%%%%%%%%%%%%%%%%%%%%%%%%  acknowledgement.tex (start)  %%%%%%%%%%%%%%%%%%%%%%

%%%%%%%%%%%%%%%%%%%%%%%%%%%%%%%%%  bibliography.tex (start)  %%%%%%%%%%%%%%%%%%%%%%%%%%%%%%

% 
%==================== bibliography===========================
% 
 
% 
%%%%%%%%%%%%%%%%%%%%%%%%%%%%%%%%%  bibliography.tex (end)  %%%%%%%%%%%%%%%%%%%%%%%%%%%%%%

\end{document}